\documentclass[]{aa}  

\usepackage{graphicx}
\usepackage{txfonts}
\usepackage{adjustbox}
\usepackage{caption}
\usepackage{float}
\usepackage[flushleft]{threeparttable}
\usepackage{pdfpages}
\usepackage{longtable}
\usepackage{textcomp}
\usepackage{multirow}
\usepackage{enumerate}
\usepackage{hhline}

\begin{document}

   \title{The 500 ks \textit{Chandra} observation of the $z=6.31$ QSO SDSS J1030+0524}

   %\subtitle{}

   \author{R. Nanni
          \inst{1,2}
          \and R. Gilli
          \inst{1}
          \and C. Vignali
          \inst{2,1}
          \and M. Mignoli
          \inst{1}
          \and A. Comastri
          \inst{1}
          \and E. Vanzella
          \inst{1}
          \and G. Zamorani
          \inst{1}
          \and F. Calura
          \inst{1}
          \and G. Lanzuisi
          \inst{2}
          \and M. Brusa
          \inst{2}
          \and P. Tozzi
          \inst{3}
           \and K. Iwasawa
          \inst{4,5}
          \and M. Cappi
          \inst{1}
          \and F. Vito
          \inst{6,7}
          \and B. Balmaverde
          \inst{8}
          \and T. Costa
          \inst{9}
          \and G. Risaliti 
          \inst{3,10}
          \and M. Paolillo
          \inst{11,12,13}
          \and I. Prandoni
          \inst{14}
          \and E. Liuzzo
          \inst{14}
          \and P. Rosati
          \inst{15}
          \and M. Chiaberge
          \inst{16,17}
          \and G. B. Caminha
          \inst{18}
          \and E. Sani
          \inst{19}
           \and N. Cappelluti
          \inst{20,21,22}
          \and C. Norman
          \inst{16,17}
          }

   \institute{INAF - Osservatorio di Astrofisica e Scienza dello Spazio di Bologna, via Gobetti 93/3 - 40129 Bologna - Italy
         \and
             Dipartimento di Astronomia, Universit\`a degli Studi di Bologna, via Gobetti 93/2, 40129 Bologna, Italy
             \and
             INAF, Osservatorio Astrofisico di Arcetri, Largo E. Fermi 5, I-50125 Firenze, Italy
             \and
             Institut de Ci\`encies del Cosmos (ICCUB), Universitat de Barcelona (IEEC-UB), Mart\'i i Franqu\`es, 1, 08028 Barcelona, Spain
             \and
             ICREA, Pg. Llu\'is Companys 23, 08010 Barcelona, Spain
             \and
             Department of Astronomy \& Astrophysics, 525 Davey Lab, The Pennsylvania State University, University Park, PA 16802, USA
             \and
             Institute for Gravitation and the Cosmos, The Pennsylvania State University, University Park, PA 16802, USA
             \and
             Scuola Normale Superiore, Piazza dei Cavalieri 7, 56126 Pisa, Italy
             \and
             Leiden Observatory, Leiden University, PO Box 9513, NL-2300 RA Leiden, the Netherlands
             \and
             Dipartimento di Fisica e Astronomia, Universit\`a di Firenze, Via G. Sansone 1, 50019 Sesto Fiorentino (FI) , Italy
             \and
             Dipartimento di Fisica "Ettore Pancini", Universit\`a di Napoli Federico II, via Cintia, I-80126 Napoli, Italy
             \and
             INFN-Unit\`a di Napoli, via Cintia 9, I-80126 Napoli, Italy
             \and
             Agenzia Spaziale Italiana - Science Data Center, Via del Politecnico snc, I-00133 Roma, Italy
             \and
             INAF-Istituto di Radioastronomia, via P. Gobetti 101, 40129 Bologna, Italy
             \and
             Universit\`a degli Studi di Ferrara, Italy
             \and
             Space Telescope Science Institute, 3700 San Martin Dr., Baltimore, MD 21210, USA
             \and
             Johns Hopkins University, 3400 N. Charles Street, Baltimore, MD 21218, USA
             \and
             Kapteyn Astronomical Institute, University of Groningen, Postbus 800, 9700 AV Groningen, The Netherlands
             \and
             European Southern Observatory, Alonso de Cordova 3107, Casilla 19, Santiago 19001, Chile
             \and
             Physics Department, University of Miami, Coral Gables, FL 33124
             \and
             Yale Center for Astronomy and Astrophysics, P.O. Box 208121, New Haven, CT 06520, USA 
             \and
             Department of Physics, Yale University, P.O. Box 208121, New Haven, CT 06520, USA 
             \\
             }

   \date{}
 
  \abstract
  {We present the results  from a $\sim500$ ks \textit{Chandra} observation of the $z=6.31$ QSO SDSS J1030+0524. This is the deepest X-ray observation to date of a $z\sim6$ QSO. The QSO is detected with a total of 125 net counts in the full ($0.5-7$ keV) band and its spectrum can be modeled by a single power-law model with photon index of $\Gamma = 1.81 \pm 0.18$ and full band flux of $f=3.95\times 10^{-15}$ erg s$^{-1}$ cm$^{-2}$. When compared with the data obtained by XMM-\textit{Newton} in 2003, our \textit{Chandra} observation in 2017 shows a harder ($\Delta \Gamma \approx -0.6$) spectrum and a 2.5 times fainter flux. Such a variation, in a timespan of $\sim2$ yrs rest-frame, is unexpected for such a luminous QSO powered by a $> 10^9 \: M_{\odot}$ black hole.
  %making SDSS J1030+0524 the first observed variable QSO at such high redshift.
The observed source hardening and weakening could be related to an intrinsic variation in the accretion rate. However, the limited photon statistics does not allow us to discriminate between an intrinsic luminosity and spectral change, and an absorption event produced by an intervening gas cloud along the line of sight.\\
We also report the discovery of diffuse X-ray emission that extends for 30"x20" southward the QSO with a signal-to-noise ratio of $\sim$6, hardness ratio of $HR=0.03_{-0.25}^{+0.20}$, and soft band flux of $f_{0.5-2 \: keV}= 1.1_{-0.3}^{+0.3} \times 10^{-15}$ erg s$^{-1}$ cm$^{-2}$, that is not associated to a group or cluster of galaxies.
We discuss two possible explanations for the extended emission, which may be either associated with the radio lobe of a nearby, foreground radio galaxy (at $z \approx 1-2$), or
ascribed to the feedback from the QSO itself acting on its surrounding environment, as proposed by simulations of early black hole formation.
%We conclude that further X-ray deep observations are needed to confirm the X-ray variability and to understand the origin of the diffuse emission.
  }
  
               
   \maketitle

%
%-------------------------------------------------------------------

\section{Introduction}

The study of high-redshift active galactic nuclei (AGN) represents one of the frontiers of modern astrophysics. In the past decades, more than 200 quasars (QSOs) with spectroscopic redshift $z > 5.5$ were discovered by wide-area optical and near-IR (NIR) surveys (\citealt{Fan06}; \citealt{Wil10}; \citealt{Ven13}; \citealt{Ban16}; \citealt{Mat16}; \citealt{Ree17}; \citealt{Tan17}; \citealt{Yan17}; \citealt{Ban17}).

Multi-wavelength observations showed that these QSOs are evolved systems with large black hole masses ($10^8-10^{10} \; M_{\odot}$; \citealt{Mor11}; \citealt{Wu15}), and large amount of gas and dust, and intense star formation in their host galaxies ($M_{gas} \sim 10^{9-10} \; M_{\odot}, M_{dust} \sim 10^{8-9} \; M_{\odot}$, SFR up to 1000 $M_{\odot}$/yr; e.g., \citealt{Cal14}; \citealt{Ven16}; \citealt{Ven17}; \citealt{Gal17a}; \citealt{Dec18}).
Optical and NIR observations showed that the broad-band spectral energy distributions (SEDs) and the rest-frame NIR/optical/UV spectra of QSOs have not significantly evolved over cosmic time (e.g., \citealt{Mor11}; \citealt{Bar15}).
Only 29 of these high-z QSOs have been studied through their X-ray emission (e.g., \citealt{Bra02}; \citealt{Far04};  \citealt{Vig05};  \citealt{She06};  \citealt{Mor14};  \citealt{Pag14};  \citealt{Ai16}; \citealt{Gal17b}). In particular, our group performed a systematic analysis of X-ray archival data of all the 29 QSOs at $z> 5.5$ observed so far with \textit{Chandra}, XMM-\textit{Newton} and \textit{Swift}-XRT, concluding that the X-ray spectral properties of high-redshift QSOs do not differ significantly from those of AGN at lower redshift (\citealt{Nan17}).

How the $10^{8-10}$ $M_{\odot}$ BHs powering $z\sim6$ QSOs could form and grow within 1 Gyr (the age of the Universe at $z\approx6$) is still a challenge for theory. Different scenarios have been proposed to explain the formation of the BH seeds that eventually became SMBHs by $z\sim6$. The two most promising ones involve either the remnants of PopIII stars (100 $M_{\odot}$; e.g., \citealt{MR01}), or more massive (10$^{4-6}$ $M_{\odot}$) BHs formed from the direct collapse of primordial gas clouds (e.g., \citealt{Vol08}; \citealt{Agar14}; see also \citealt{Val16} for a model of seed formation at different mass scales). In the case of low-to-intermediate mass ($M\le10^4$) seeds, super-Eddington accretion is required to form the black-hole masses observed at $z > 6$ (e.g., \citealt{Mad14}; \citealt{Vol16}; \citealt{Pez17}).

There is general agreement that early massive BHs form in overdense environments, that may extend up to 10 physical Mpc (pMpc), and host large gas quantities (\citealt{Over09}; \citealt{Dub13}; \citealt{Cos14}; \citealt{Bar17}). According to simulations, the fields around high-redshift QSOs are expected to show galaxy overdensities, which probably represent the progenitors of the most massive clusters in the local Universe (\citealt{Spr05}). In the past decade, large efforts have been made to find overdense regions in fields as large as 2x2 pMpc around $z\sim6$ QSOs (e.g., \citealt{Sti05}; \citealt{Hus13}; \citealt{Ban13}; \citealt{Simp14}; \citealt{Mazz17}), but the results were inconclusive. Some of them ascribed the lack of detection of overdensities at very high-z to the strong ionizing radiation from the QSO that may prevent star formation in its vicinity. The presence of strong gas jets and/or radiation feedback extending up to few hundreds of kpc at $z=6$ is, in fact, predicted in modern simulations of BHs formation (\citealt{Cos14}; \citealt{Bar17}).

The QSO SDSS J1030+0525 at $z=6.31$ (\citealt{Fan01}) was one of the first $z\sim6$ QSOs discovered by the Sloan Digital Sky Survey (SDSS), and its field is part of the Multiwavelength Chile-Yale survey (MUSYC). It has also been covered by HST/WFC3.
Near-IR spectroscopy showed that it is powered by a BH with mass of $1.4\times 10^9$ $M_{\odot}$ (\citealt{Kur07}; \citealt{DeR11}). It was not detected in the submillimeter (\citealt{Pri03}) and radio bands (\citealt{Pet03}), but it was detected in the X-rays by \textit{Chandra} (one 8-ks snapshot in 2002; \citealt{Bra02}) and by XMM-\textit{Newton} (one 105-ks observation in 2003; \citealt{Far04}). In concordance with literature results on other $z\sim6$ QSOs, the rest-frame optical continuum shape and luminosity of this QSO are consistent with those of lower redshift AGN (\citealt{Fan01}). The X-ray spectrum is instead possibly steeper than standard QSOs spectra ($\Gamma \sim 2.1-2.4$; \citealt{Far04} and \citealt{Nan17}). Deep and wide imaging observations of a $8\times8$ pMpc$^2$ region around SDSS J1030+0524 also showed that this field features the best evidence to date of an overdense region around a $z\sim6$ QSO (\citealt{Mors14}; \citealt{Bal17}).
In the last few years, our group has obtained data in the optical and X-ray bands to further investigate and confirm the presence of the putative overdensity, and to obtain one of the highest quality spectrum ever achieved in X-ray for a QSO at $z\sim6$. In particular, we report here the results from our $\sim$500 ks \textit{Chandra} ACIS-I observation of SDSS J1030+0524 that represents the deepest X-ray look at a $z > 6$ QSO to date.

The paper is organized as follows. In \S 2 we describe the X-ray \textit{Chandra} data, and the data reduction procedure. In \S 3 we report the data analysis and spectral fitting, the X-ray variability, and the study of the diffuse emission around the QSO. In \S 4 we discuss the physical conditions that can be responsible for the X-ray observed features, provide the multi-band SED of the QSO, and discuss the possible origins of the diffuse emission. In \S 5 we give a summary of our results.
Throughout this paper we assume $H_0 = 70$ km s$^{-1}$ Mpc$^{-1}$, $\Omega_{\Lambda} = 0.7$, and $\Omega_M = 0.3$ (\citealt{Ben13}), and errors are reported at 68\% confidence level if not specified otherwise. Upper limits are reported at the 3$\sigma$ confidence level. 
%-------------------------------------------------------------------

\section{\textit{Chandra} observations}

SDSS J1030+0524 was observed by \textit{Chandra} with ten different pointings between January and May 2017 for a total exposure of 479 ks.  Observations were taken using the Advanced CCD Imaging Spectrometer (ACIS) instrument and the target was positioned on the ACIS-I3 chip, at roll-angle $\sim$64\textdegree$\!$ for the first 5 observations, and at roll-angle $\sim$259\textdegree, for the others. 
The ten observations (hereafter ObsIDs) cover a total area of roughly 335 arcmin$^2$ in size and the exposure times of the individual observations range from 26.7 to 126.4 ks. A summary of the observational parameters is provided in Table 1.
The data were reprocessed using the \textit{Chandra} software CIAO v. 4.8 using the \textit{vfaint} mode for the event telemetry format. Data analysis was carried out using only the events with ASCA grades 0, 2, 3, 4 and 6. We then produced X-ray images in the soft ($0.5-2$ keV), hard ($2-7$ keV) and full ($0.5-7$ keV) bands for each ObsID.

After this basic reduction, we corrected the astrometry (applying shift and rotation corrections) of the individual ObsIDs using as reference catalog the WIRCAM catalog comprising 14777 J-band selected sources down to $J_{AB}=24.5$ (\citealt{Bal17}). First we created exposure maps and psf maps for all ObsIDs using the CIAO tools \textit{fluximage} and \textit{mkpsfmap}, respectively. The exposure and psf maps were computed in the full band at the 90\% of the encircled energy fraction (EEF) and at an energy of 1.4 keV.
Then, we ran the \textit{Chandra} source detection task \textit{wavdetect} on the $0.5-7$ keV images to detect sources to be matched with the J-band detected objects.
We set the detection threshold to 10$^{-6}$ and wavelet scales up to 8 pixels in order to get only the brightest sources with a well defined X-ray centroid and we also provided the exposure and psf maps.
For the match we considered only the X-ray sources with a positional error\footnote{Computed as: $\sqrt{\sigma_{RA}^2+\sigma_{Dec}^2}$, where $\sigma_{RA}$ and $\sigma_{Dec}$ are the errors on Right Ascension and Declination, respectively.} below $\sim$0.4", in order to avoid sources with too uncertain centroid position. We used the CIAO tool \textit{wcs\_match} and \textit{wcs\_update} to match the sources and correct the astrometry, and create new aspect solution files. We considered a matching radius of 2" and we applied both translation and rotation corrections. The new aspect solutions were then applied to the event files and the detection algorithm was run again (using the same \textit{wavdetect} parameters and criteria adopted previously). The applied astrometric correction reduces the mean angular distance between the X-ray sources and their J-band counterparts from 
$\theta=0.253$" to $\theta=0.064$".
Finally, we stacked the corrected event files using the \textit{reproject\_obs} task and derived a new image of the field. In Figure 1 we display the final \textit{Chandra} full band image around the QSO position.
\begin{figure}
 \centering
 \includegraphics[height=9.5cm, width=9.5cm, keepaspectratio]{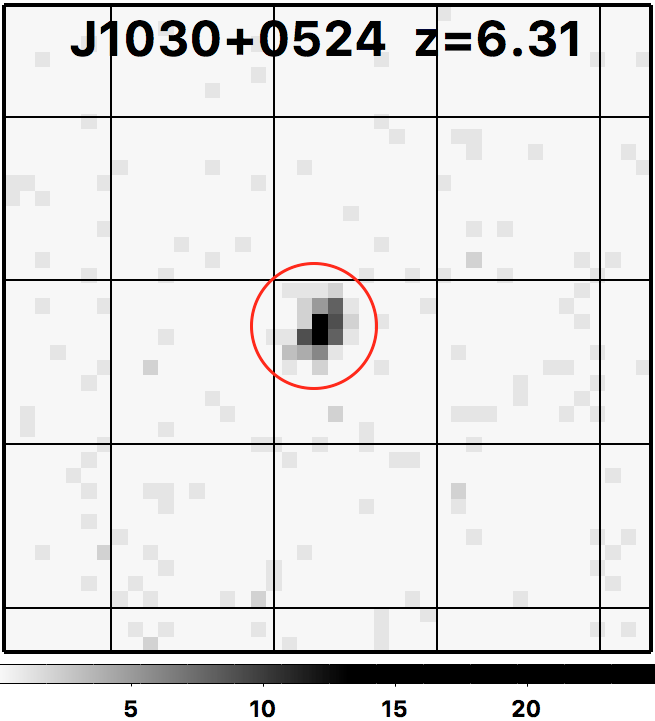}
 \caption{Full-band ($0.5-7$ keV) \textit{Chandra} ACIS-I image of SDSS J1030+0524. The red circle represents our extraction region (2" radius). The grid separation is 5" and the cutout spans 20"x20" on the sky. Units on the colorbar are counts per pixel.}
\end{figure}

\section{Results}

\subsection{Timing analysis}

The long total exposure taken on a time span of five months enabled us to study the possible presence of flux and spectral variability.
We extracted the number of counts in each ObsID from circular regions centered at the optical position of the QSO. We used a radius of 2", corresponding to 95\% of the encircled energy fraction (EEF) at 1.5 keV, for the source extraction, and a nearby region (free of serendipitously detected sources), with a 100 times larger area, for the background extraction. In the final four columns of Table 1 we report the full ($0.5-7$ keV), soft ($0.5-2$ keV), hard ($2-7$ keV) band net counts, extracted in each single observation, and the hardness ratios (HRs), computed as HR $= \frac{H-S}{H+S}$ where H and S are the net counts in the hard and soft bands, respectively. 

We first determined whether the QSO varied during the \textit{Chandra} observations by applying a $\chi^2$ test to its entire light curve in the full band. This is computed as
\begin{equation}
	\chi^2_{\nu} = \frac{1}{N-1}\sum_{i=1}^N \frac{(f_i-\bar{f})^2}{\sigma_i^2}
\end{equation}
where $f_i$ and $\sigma_i$ are the full band count rates and its error in the $i$th observation, $\bar{f}$ is the average count rate of the source and $N$ is the number of the X-ray observations.
The null hypothesis is that the count rate in each epoch is consistent with the mean count rate of the entire light curve, within the errors.
We show the distribution of the count rates in the three bands (full, soft, and hard) vs time, starting from the first observation, in Figure 2, where the red lines represent the mean weighted value of the rates.
We computed the probability by which the null hypothesis can be rejected ($p$), and obtained $p\sim0.47$ (0.44, and 0.40) for the full (soft, and hard) band, respectively. We then conclude that there is no evidence of count rate variability among our \textit{Chandra} observations.
\begin{figure}
 \includegraphics[height=15cm, width=8cm, keepaspectratio]{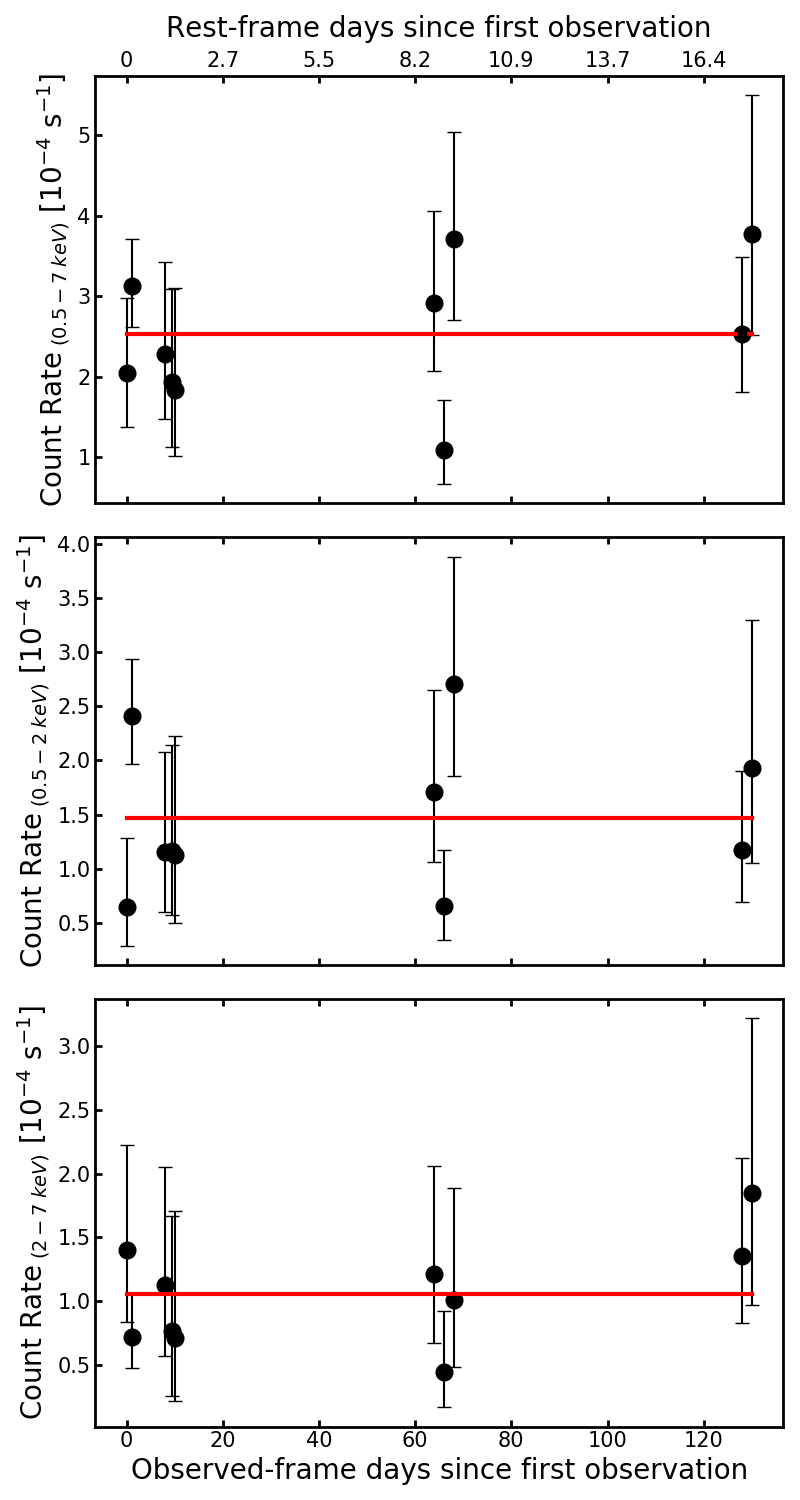}
 \caption{Count rate of SDSS J1030+0524 in the three X-ray bands (full in the top, soft in the middle, and hard in the bottom panel) extracted from the ten \textit{Chandra} observations vs the days since the first observation. Errors are reported at the 1$\sigma$ level. The red solid lines represent the weighted mean.}
\end{figure}
The HR distribution vs the observation time is reported in Figure 3.
\begin{figure}
 \includegraphics[height=7cm, width=9cm, keepaspectratio]{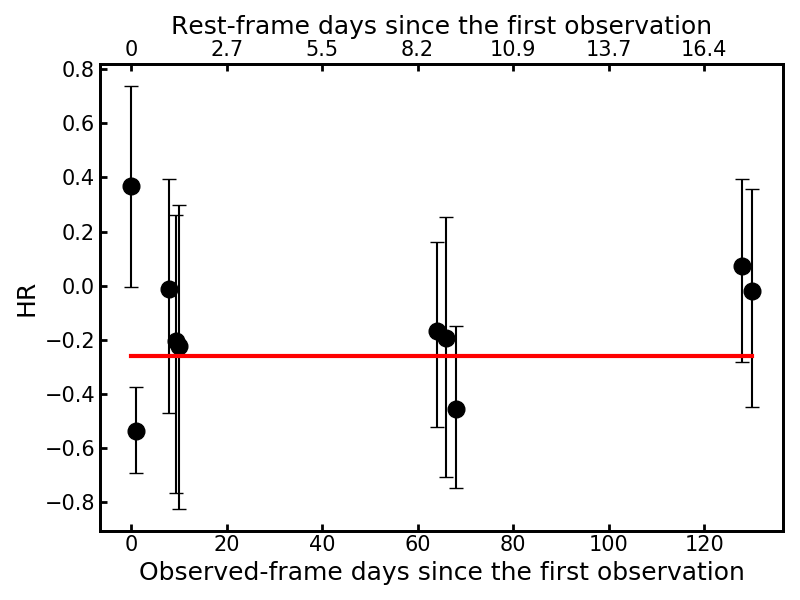}
 \caption{Hardness-ratio of SDSS J1030+ 0524 in the ten \textit{Chandra} observations vs the days since the first observation. Errors are reported at the 1$\sigma$ level. The red solid line represents the weighted mean.}
\end{figure}
Despite some fluctuations, also the HRs of the different observations show no significant variability ($p=0.53$). 

\begin{table*}
  \centering
  \captionsetup{justification=centering, labelsep = newline}
      \caption[]{Data information on J1030+0524}
      \begin{adjustbox}{center, max width=\textwidth}
         \begin{tabular}{c c c c c c c c}
            \hline
            \hline \rule[0.7mm]{0mm}{3.5mm}
            ObsID & Date & $\theta^{a}$ & $t_{exp}^{b}$ & Cts$_{(0.5-7 \: keV)}^c$ & Cts$_{(0.5-2 \: keV)}^c$ & Cts$_{(2-7 \: keV)}^{c}$ & HR$^d$\\
            & & [\textdegree] & [ks] & & & & \\
%            (1) & (2) & (3) & (4) & (5) & (6) & (7) & (8) & (9)\\
            \hline \rule[0.7mm]{0mm}{3.5mm}
            18185 & 2017 Jan 17 & 64.2 & 46.3  & 9.5$_{-3.1}^{+4.3}$ & 3.0$_{-1.7}^{+3.0}$ & 6.5$_{-2.6}^{+3.8}$ & +0.37$_{-0.37}^{+0.38}$ \\
            \rule[0.7mm]{0mm}{3.5mm}
            19987 & 2017 Jan 18 & 64.2 & 126.4  & 39.5$_{-6.4}^{+7.5}$ & 30.4$_{-5.6}^{+6.7}$ & 9.1$_{-3.1}^{+4.3}$ & -0.54$_{-0.17}^{+0.15}$\\   
            \rule[0.7mm]{0mm}{3.5mm}
 	   18186 & 2017 Jan 25 & 64.2 & 34.6  & 7.9$_{-2.8}^{+4.0}$ & 4.0$_{-1.9}^{+3.2}$ & 3.9$_{-1.9}^{+3.2}$ & -0.01$_{-0.41}^{+0.46}$\\   
	   \rule[0.7mm]{0mm}{3.5mm}
  	   19994 & 2017 Jan 27 & 64.2 & 32.7 & 6.3$_{-2.6}^{+3.8}$ & 3.8$_{-1.9}^{+3.2}$ & 2.5$_{-1.7}^{+2.9}$ &   -0.21$_{-0.47}^{+0.56}$\\  
	    \rule[0.7mm]{0mm}{3.5mm}
  	   19995 & 2017 Jan 27 & 64.2 & 26.7 & 4.9$_{-2.2}^{+3.4}$ & 3.0$_{-1.7}^{+3.0}$ & 1.9$_{-1.3}^{+2.7}$  &  -0.22$_{-0.52}^{+0.60}$\\ 
	   \rule[0.7mm]{0mm}{3.5mm}
  	   18187 & 2017 Mar 22 & 259.2 & 40.4 & 11.8$_{-3.4}^{+4.6}$ & 6.9$_{-2.6}^{+3.8}$ & 4.9$_{-2.2}^{+3.4}$ &  -0.17$_{-0.33}^{+0.35}$\\ 
	   \rule[0.7mm]{0mm}{3.5mm}
 	   20045 & 2017 Mar 24 & 259.2 & 61.3 & 6.7$_{-2.6}^{+3.8}$ & 4.0$_{-1.9}^{+3.2}$ & 2.7$_{-1.7}^{+3.0}$ &  -0.19$_{-0.45}^{+0.51}$\\
	   \rule[0.7mm]{0mm}{3.5mm}
  	   20046 & 2017 Mar 26 & 259.2 & 36.6 & 13.6$_{-3.7}^{+4.9}$ & 9.9$_{-3.1}^{+4.3}$ & 3.7$_{-1.9}^{+3.2}$ &  -0.46$_{-0.31}^{+0.29}$\\
	   \rule[0.7mm]{0mm}{3.5mm}
	   19926 & 2017 May 25 & 262.2 & 49.4 &12.5$_{-3.6}^{+4.7}$ & 5.8$_{-2.4}^{+3.6}$ & 6.7$_{-2.6}^{+3.8}$ &  +0.07$_{-0.32}^{+0.35}$\\
	   \rule[0.7mm]{0mm}{3.5mm}
 	   20081 & 2017 May 27 & 262.2 & 24.9 & 9.4$_{-3.1}^{+4.3}$ & 4.8$_{-2.2}^{+3.4}$ & 4.6$_{-2.2}^{+3.4}$ &  -0.02$_{-0.38}^{+0.43}$\\[2pt]
            \hline 
         \end{tabular}
        \end{adjustbox}
        \begin{tablenotes}
        		\footnotesize
		\item \begin{enumerate}[(a)]
			\item Roll-angle in degrees of the ACIS-I instrument.
			\item Exposure time after background flare removal.
			\item Net counts in the full ($0.5-7$ keV), soft ($0.5-2$ keV), and hard ($2-7$ keV) bands, respectively. Errors on the X-ray counts were computed according to Table 1 and 2 of 					\citet{Geh86} and correspond to the 1$\sigma$ level in Gaussian statistics.
			\item The hardness ratio is defined as $HR = \frac{H-S}{H+S}$ where H and S are the counts in the hard (2.0-7.0 keV) and soft ($0.5-2$ keV) bands.
			We calculated errors at the 1$\sigma$ level for the hardness ratio following the method described in \S 1.7.3 of Lyons (1991).
		\end{enumerate}
	 \end{tablenotes}
   \end{table*}   

%-------------------------------------- 

\subsection{Spectral analysis}

The lack of significant flux and hardness ratio variability allowed us to combine the ten spectra together (each extracted from the corresponding event file), and obtain a final spectrum with 125 net counts in the full band. The spectral channels were binned to ensure a minimum of one count for each bin, and the best-fit model was decided using the Cash statistics (\citealt{Cas79}).
First, we modeled the spectrum with a simple power-law, using XSPEC v. 12.9 (\citealt{Arn96}), with a Galactic absorption component fixed to $2.6 \times 10^{20}$ cm$^{-2}$ (the value along the line of sight towards the QSO, \citealt{Kal05}). We found that the best-fit photon index is $\Gamma=1.81_{-0.18}^{+0.18}$ (C-stat = 88.3 for 93 d.o.f.), and the flux in the $0.5-2$ keV band is 1.74$_{-0.38}^{+0.11}\times 10^{-15}$ erg s$^{-1}$ cm$^{-2}$.
The value of the photon index is consistent with the mean photon indices obtained by jointly fitting spectra of unobscured QSOs at the same and at lower redshifts ($\Gamma \sim 1.6-2.0$ for $1 \le z \le 7$; e.g., \citealt{Vig05}; \citealt{She06}; \citealt{Jus07}; \citealt{Nan17}; \citealt{Vit17}), but it is flatter than the XMM-\textit{Newton} value found for the same QSO by \citet{Far04} ($\Gamma=2.12_{-0.11}^{+0.11}$) and by \citet{Nan17} ($\Gamma=2.39_{-0.30}^{+0.34}$; although they fit a power-law plus intrinsic absorption model). Also our measured soft flux is 3.6 times lower than the one derived by \citet{Far04} ($f_{0.5-2}=6.3\times 10^{-15}$ erg s$^{-1}$ cm$^{-2}$), and this difference is significant at the 4$\sigma$ level (see \S 3.3 for a detailed study of the long-term variability). The spectrum and its best-fit model and residuals are shown in Figure 4.
\begin{figure}
 \includegraphics[height=15cm, width=9cm, keepaspectratio]{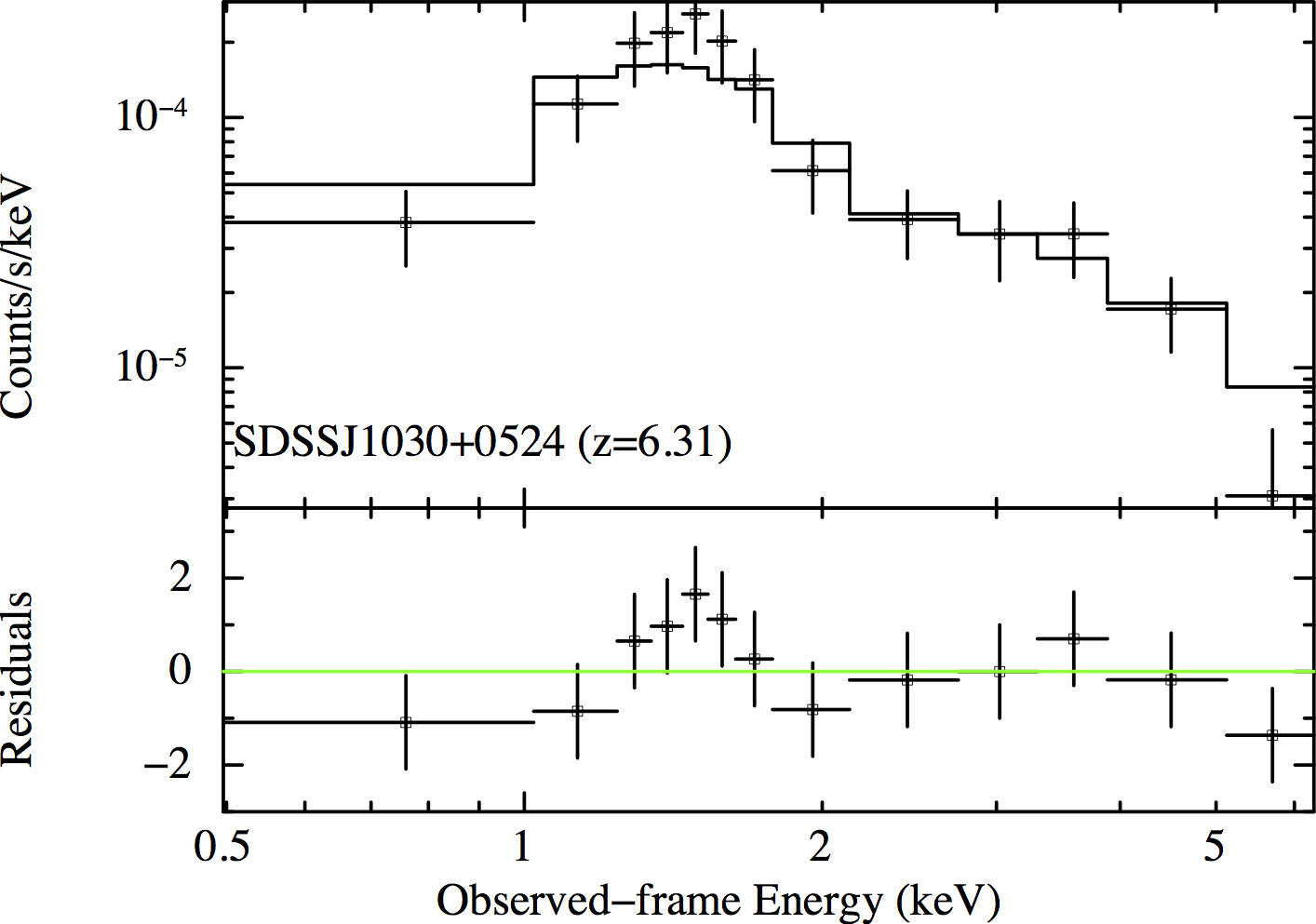}
 \caption{X-ray spectrum of SDSS J1030+0524 fitted with a power-law model ($\Gamma=1.81_{-0.18}^{+0.18}$). In the bottom panel we report the residuals [(data$-$model)/error]. For display purposes we adopted a minimum binning of ten counts per bin.}
\end{figure}

We performed other fits by adding spectral components to the power-law plus Galactic absorption model.
First, we added an intrinsic absorption component. Because of the very high redshift of the QSO, this fit is sensitive only to very high values of obscuration (N$_H \ge 10^{23-24}$ cm$^{-2}$). We found that the column density is poorly constrained and consistent with no absorption (N$_H = 4.6_{-4.6}^{+2.7}\times 10^{23}$ cm$^{-2}$), as it may be expected for a luminous Type 1 QSO such as J1030+0524. Then, we fit the same model with the photon index fixed to $\Gamma=2.39$ (best-fit value found in the XMM-\textit{Newton} data for this QSO by \citealt{Nan17}) and we found N$_H = 5.3_{-1.7}^{+1.8}\times 10^{23}$ cm$^{-2}$.
To search for the presence and significance of a narrow emission iron line, we also added to the power-law model (with photon index fixed to $\Gamma=1.8$) a Gaussian line, with rest-frame energy of 6.4 keV and width of 10 eV (both fixed in the fit).
We obtained a fit with similar quality (C-stat/d.o.f. = 88.4/93) to that of the single power-law model and we derived an upper limit for the rest-frame iron line equivalent width of EW $\le$ 460 eV. We also checked the presence of iron lines at rest-frame energy of 6.7 and 6.9 keV (as expected from highly ionized iron, FeXXV and FeXXVI), obtaining a rest-frame equivalent width of EW $\le$ 420 eV, in both cases.
Considering that we are sampling rest frame energies in the range 3.5-50 keV, where a hardening of AGN spectra is often observed because of the so called "Compton-reflection hump"\footnote{The "Compton-reflection hump" is radiation from the hot corona that is reprocessed by the accretion disk, and peaks at $\sim$30 keV.}, we checked the possible contribution to the spectrum of a reflected component (\textit{pexrav} model) finding that the photon index is poorly constrained ($\Gamma = 1.72_{-0.16}^{+0.95}$).
The possible presence of a FeXXVI emission line at 6.7 keV in the XMM-\textit{Newton} spectrum, with significance at the 2.5$\sigma$ level, suggested us to fit the \textit{Chandra} spectrum with a power-law model plus a reflection ionized component (\textit{reflionx}). However, we obtained an ionization parameter that is poorly constrained and so is the normalization of the reflection component. 

\begin{figure*}
\begin{center}
 \includegraphics[height=8cm, width=15cm, keepaspectratio]{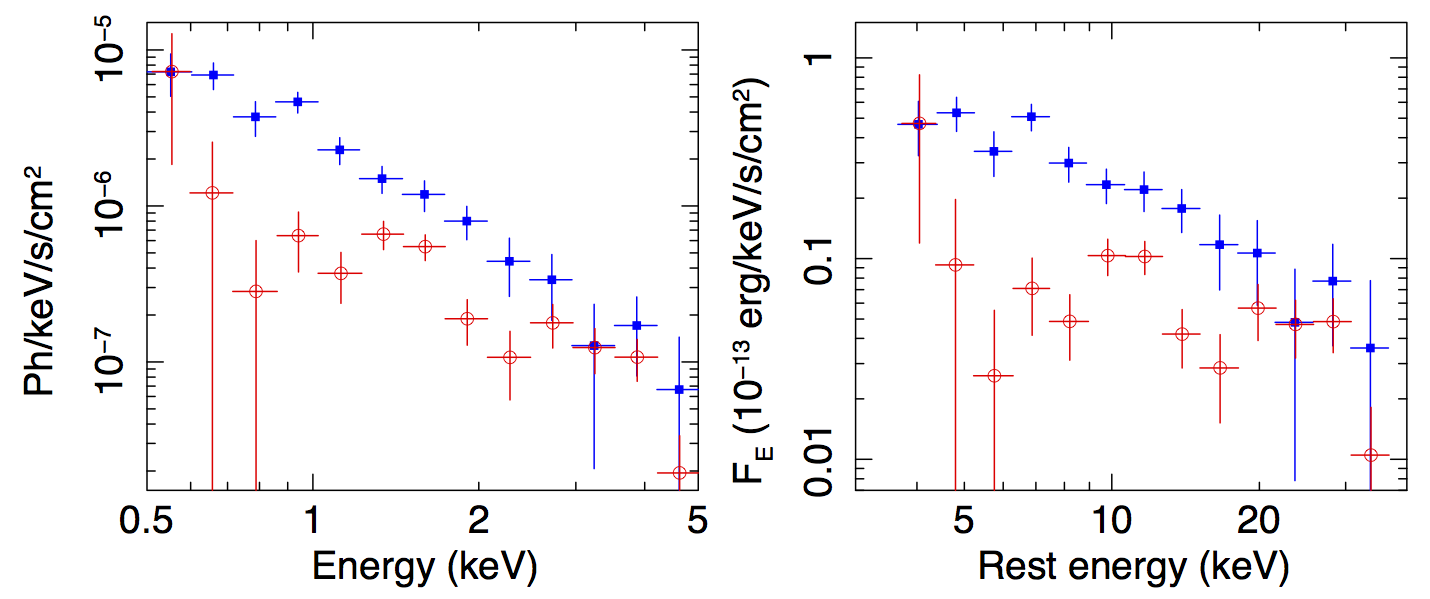}
 \caption{Comparison between the 2003 XMM-\textit{Newton} (in blue) and 2017 \textit{Chandra} (in red) spectrum of SDSS J1030+0524. Both spectra have been corrected for the effective response. The XMM-\textit{Newton} spectrum is the average of the three EPIC cameras (with SNR weighting). The left panel shows the observed-frame spectra, while the right panel shows the rest-frame one.}
 \end{center}
\end{figure*}

%(adding a \textit{reflionx} component to our power-law model). We found a steep photon index of $\Gamma = 2.48_{-0.56}^{+0.30}$ (C-stat/d.o.f. = 87.0/91) with a low level of ionization ($X_i \le 200$ erg cm s$^{-1}$). However, such a model should also dominate the emission seen at observed-frame $\sim$3 keV in the XMM-\textit{Newton} spectrum, with a consequently steepening of the power-law slope found with XMM-\textit{Newton} ($\Gamma = 2.7 \pm 0.2$) that is too high to be considered physically acceptable.

Finally, we noted the possible presence of a dip in the \textit{Chandra} spectrum at $\sim$2.4 keV observed-frame ($\sim$17.5 keV rest-frame; see red spectrum in Figure 5).
Previous studies of X-ray AGN spectra revealed the presence of blue-shifted Fe K-shell absorption lines, at rest-frame energies > 7 keV, possibly related to ultra-fast outflows (UFOs) of gas ejected from the QSOs with velocities $\ge \: 10^4$ km s$^{-1}$ (e.g., \citealt{Tomb13}; \citealt{Tomb15}). We then checked the presence of absorption features producing this dip, fitting the spectrum with a power-law model (with $\Gamma$ free to vary) plus an absorption line component. We generated 10$^4$ fake spectra (using the same response files and statistics of our original spectrum), and fit them with the power-law plus absorption line model; the adopted procedure is fully described in \citet{Lan13b} and \citet{Tomb13}. Comparing the C-stat distribution of the 10$^4$ fake spectra with the one obtained for the original one, we found that the absorption feature is not significant (<2$\sigma$ level).
We also noted the presence of a dip in the 5-10 keV rest-frame energy range (Figure 5), followed by a rise at lower energies, that could be related to the absorption by warm absorbers; however, this rise at low energies contains only 1-2 counts per bin.
We tried to fit this low-energy dip with a warm absorber ($warmabs$\footnote{$Warmabs$ can be used within XSPEC to fit to observed spectra the results of XSTAR, a software package for calculating the physical conditions and emission spectra of photoionized gas (\citealt{K&B01}).}) plus a power-law model, fixing the photon index to the \textit{Chandra} ($\Gamma=1.81$) and XMM-\textit{Newton} ($\Gamma=2.39$) best-fit values. 
In both cases we found best-fit values of column density (N$_H \sim 6\times 10^{23}$ cm$^{-2}$) and ionization parameter ($log(x_i) \sim 2$) that point back to a cold absorber scenario (the one we tested with the power-law plus absorption model). Furthermore, these values are not well constrained due to the limited counting statistics. Finally, we also tried to fit the rise at low energies with a power-law plus a partial covering absorption model ($zxipcf$), fixing again the photon index to the \textit{Chandra} ($\Gamma=1.81$) and XMM-\textit{Newton} ($\Gamma=2.39$) best-fit values. Also in that case, the result points back to a cold absorber scenario (N$_H \sim 7\times 10^{23}$ cm$^{-2}$, $log(x_i) \le 2$, and covering factor of $f\sim0.9$) with a similar statistical quality of the fit.
In Table 2 we summarize the results of our spectral analysis.

\begin{table*}
  \centering
  \captionsetup{justification=centering, labelsep = newline}
      \caption[]{Best-fit results of the \textit{Chandra} data}
      \begin{adjustbox}{center, max width=\textwidth}
         \begin{tabular}{c c c c c c c}
            \hline
            \hline \rule[0.7mm]{0mm}{3.5mm}
            Model & C-stat/d.o.f. & $\Gamma$ & Parameter & f$_{(0.5-2 \: keV)}$ & f$_{(2-7 \: keV)}$ & L$_{(2-8 \: keV)}^{rest}$\\
           (1) & (2) & (3) & (4) & (5) & (6) & (7)\\
            \hline \rule[0.7mm]{0mm}{3.5mm}
            Power-law & 88.3/93 & 1.81$_{-0.18}^{+0.18}$  & ... & 1.74$_{-0.38}^{+0.11}$ & 2.20$_{-0.55}^{+0.17}$ & 6.14$_{-2.21}^{+0.85}$  \\
            \rule[0.7mm]{0mm}{3.5mm}
            Power-law plus absorption & 88.2/92 & 1.87$_{-0.21}^{+0.48}$  & N$_H = 4.6_{-4.6}^{+2.7}\times10^{23}$ cm$^{-2}$& 1.68$_{-0.23}^{+0.13}$ & 2.18$_{-0.27}^{+0.19}$ & 6.97$_{-1.55}^{+1.69}$ \\   
            \rule[0.7mm]{0mm}{3.5mm}
            Power-law plus absorption$^{\diamond}$ & 89.4/93 & 2.39  & N$_H = 5.3_{-1.7}^{+1.8}\times10^{23}$ cm$^{-2}$& 1.42$_{-0.22}^{+0.23}$ & 2.05$_{-0.29}^{+0.28}$ & 17.2$_{-3.7}^{+4.3}$ \\   
            \rule[0.7mm]{0mm}{3.5mm}
 	   Power-law plus iron line$^{\ddagger}$ & 88.4/93 & 1.8 & EW $\le$ 464 eV & 1.83$_{-0.11}^{+0.17}$ & 2.07$_{-0.20}^{+0.10}$ & 6.34$_{-0.38}^{+0.97}$ \\   
	   \rule[0.7mm]{0mm}{3.5mm}
  	   Power-law plus reflection & 87.9/92 & 1.72$_{-0.16}^{+0.95}$ & Rel$_{refl}$ $\le$ 14 & 1.74$_{-0.13}^{+0.12}$ & 2.20$_{-0.79}^{+5.67}$ & 5.88$_{-1.84}^{+4.82}$ \\[2pt]
            \hline 
          \end{tabular}
        \end{adjustbox}
        \begin{tablenotes}\footnotesize
		\item (1) Model fitted to the X-ray spectrum. (2) Value of the C-stat vs the degrees of freedom. (3) Photon index found or used in the fit. (4) Best-fit value of the corresponding fit-model parameter. (5), (6) Fluxes in the observed $0.5-2$ and $2-7$ keV bands in units of 10$^{-15}$ erg cm$^{-2}$ s$^{-1}$. (7) Intrinsic luminosity in the rest-frame $2-10$ keV band in units of 10$^{44}$ erg s$^{-1}$. Errors are reported at the 1$\sigma$ level and upper limits at the 3$\sigma$ level.
		\item $\diamond$ Fitting model in which $\Gamma$ was fixed to the best-fit value found in the XMM-\textit{Newton} data for this QSO by \citealt{Nan17}.
		\item $\ddagger$ For this model we report results for the case with a K$\alpha$ emission iron line with fixed rest-frame energy of 6.4 keV and width of 0.01 keV. $\Gamma$ was fixed to 1.8.
	 \end{tablenotes}
\end{table*}   
   
\subsection{Comparison with previous analysis}
      
J1030+0524 has been observed in the X-rays twice in the past: by \textit{Chandra} in 2002 and by XMM-\textit{Newton} in 2003.
As reported in \S 2.2, our derived soft band \textit{Chandra} flux is $\sim$3.6 times lower than that observed by XMM-\textit{Newton}.

We derived the observed-frame full band ($0.5-7$ keV) fluxes for the 2002, 2003, and 2017 observations to build the long-term X-ray light curve (see Figure 6).
From the fit we performed on the 2017 \textit{Chandra} observation, using a simple power-law model (first row in Table 2), we found $f_{0.5-7\:keV}=3.96_{-0.83}^{+0.18}\times 10^{-15}$ erg s$^{-1}$ cm$^{-2}$.

For the \textit{Chandra} snapshot we extracted the number of counts from a circular region with 2" radius, centered on the source position, and the background counts from a nearby circular region with ten times larger area. The source is detected with $\sim$6 net counts in the full band. Assuming a power-law with $\Gamma=1.8$, we derived a full band flux of $f_{0.5-7\:keV} = 5.4_{-2.1}^{+3.0}\times 10^{-15}$ erg s$^{-1}$ cm$^{-2}$, which is 1.4 times higher than the value found for the 2017 observation but consistent with it within the uncertainties.

For the XMM-\textit{Newton} observation, we extracted the three spectra (pn, MOS1, MOS2) from circular regions centered at the optical position of the QSOs with radius of 15", corresponding to 65\% of EEF at 1.5 keV, to avoid contamination from nearby luminous sources, while the background was extracted from a nearby region with radius of 30". We used a grouping of one count for each bin for all spectra of
the three cameras, and fit the three EPIC spectra (pn, MOS1 and MOS2) with a simple power-law model with photon index free to vary. We obtained a best-fit value $\Gamma=2.37_{-0.15}^{+0.16}$, that is consistent with the one found by \citet{Far04} but is inconsistent at the $2.4\sigma$ level with the value reported in the first column of Table 2 ($\Gamma = 1.81 \pm 0.18$), and a flux $f_{0.5-7\:keV}=9.78_{-1.18}^{+0.44}\times 10^{-15}$ erg s$^{-1}$ cm$^{-2}$, that is 2.5 times higher than the full flux derived from the longest \textit{Chandra} observation ($f_{0.5-7\:keV}=3.96_{-0.83}^{+0.18}\times 10^{-15}$ erg s$^{-1}$ cm$^{-2}$); the difference is significant at the 4.9$\sigma$ level. In Figure 6 we show the $0.5-7$ keV light curve of the QSO with the fluxes obtained from the three epochs, while in Figure 5 we show the observed-frame (left) and rest-frame (right) spectra of our \textit{Chandra} (in red) and XMM-\textit{Newton} (in blue) analyses.
We determined whether the QSO could be considered variable by applying the $\chi^2$ test to its entire light curve in the full band on year timescale, considering the first \textit{Chandra} (2002) and the XMM-\textit{Newton} (2003) observations and our longer \textit{Chandra} observation (2017).
We found from our $\chi^2$ test that the QSO has varied ($p \sim 0.99$) with a $\chi^2$ value of 8.51 (d.o.f. = 2).
\begin{figure}
 \includegraphics[height=8cm, width=9cm, keepaspectratio]{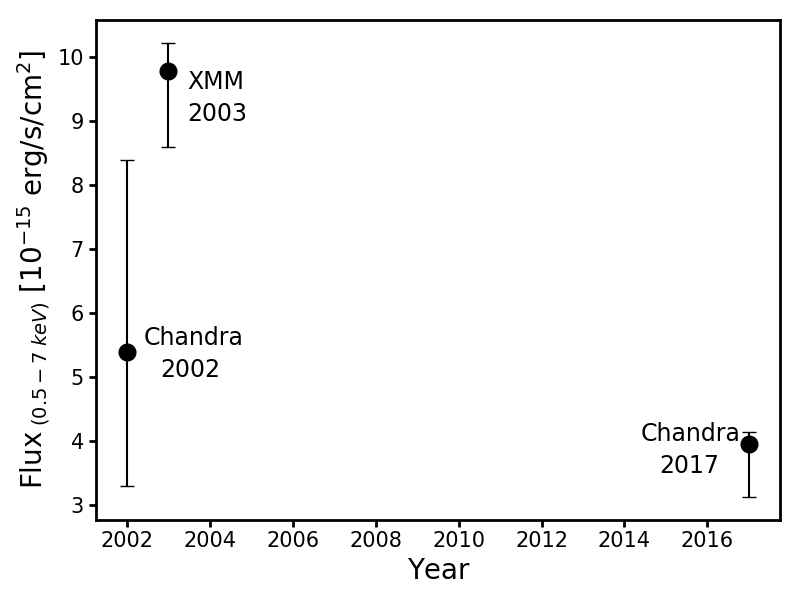}
 \caption{Long-term X-ray light curve of SDSS J1030+0524 in the $0.5-7$ keV band. Errors are reported at the 1$\sigma$ level.}
\end{figure}

Previous works (see Appendix B of \citealt{Lan13}) showed that the XMM-\textit{Newton} source spectra tend to be fitted with softer power-laws (up to 20\% difference in photon index) than those observed by \textit{Chandra}. This difference may be now possibly exacerbated by the rapid degradation of the Chandra ACIS-I effective area, which, for instance, has decreased by 18\% at 1.5 keV and by 38\% at 1 keV between the observations described in \citet{Lan13} and ours.
In order to verify whether the flux and slope variations are due to the AGN variability and not to any instrumental effect related to the different responses of \textit{Chandra} and XMM-\textit{Newton}, we performed additional checks on the XMM-\textit{Newton} and \textit{Chandra} data-sets. 
First, we changed the QSO spectral extraction parameters (e.g., size and position of both source and background extraction regions), and the QSO light curve filtering (e.g., cutting the XMM-\textit{Newton} background fluctuations adopting a different thresholds during the data reprocessing) to verify the fit stability, and found that the new best-fit parameters were fully consistent with the XMM-\textit{Newton} values reported above. 
Secondly, we selected five QSOs detected by both \textit{Chandra} and XMM-\textit{Newton} (with similar counting statistics of our QSO and observed in the central region of the data-sets), and extracted their spectra with the same extraction parameters that we adopted for J1030+0524. We found that the XMM-\textit{Newton} spectra are neither systematically steeper nor brighter than the corresponding \textit{Chandra} spectra, at least in the photon counting statistics regime considered here.
Furthermore, our normalized difference in photon index ($(\Gamma_{XMM-\textit{Newton}}/\Gamma_{Chandra})-1=0.31\pm0.04$) is three times higher than ($\sim$4$\sigma$ off) the mean value found for X-ray sources detected with similar statistic ($(\Gamma_{XMM-\textit{Newton}}/\Gamma_{Chandra})-1$=0.1 in \citealt{Lan13}).
We conclude that the XMM-\textit{Newton} results are stable and that the observed spectral variability in SDSS J1030+0524 is real. 
%We refer to \S 4.1 for a complete discussion onto the physical processes responsible for the X-ray variability.

\subsection{Diffuse emission southward the QSO}

By visual inspection of the 2017 \textit{Chandra} observation, we noted an excess of photons extending up to 25" southward of the QSO. This excess becomes more evident by smoothing the image with the task \textit{csmooth}, using a minimal (maximal) signal-to-noise ratio (SNR) of 2 (50), as shown in Figure 7 (left). This diffuse emission lies in a region in which our observations are very sensitive, as shown in Figure 7 (central).
We performed photometry on the un-smoothed image, extracting the diffuse counts and spectrum from a region with an area of 460 arcsec$^2$, shown in Figure 7 (green polygon in the central panel), and the background counts from nearby circular regions (free of X-ray point like sources)
with a total area $\sim$3 times larger. We found that the diffuse emission is highly significant, with 90 net counts, corresponding to a SNR = 5.9, and a hardness ratio HR = $0.03_{-0.25}^{+0.20}$. A hint of this diffuse emission is also visible in the XMM-\textit{Newton} observation (right panel of Figure 7), at the same sky coordinates, although its significance is less clear as it is difficult to disentangle the diffuse emission from the emission of the nearby QSO, due to the limited XMM-\textit{Newton} angular resolution. Visual inspection of Figure 7 (left) suggests that the diffuse emission may be structured into a few blobs. 
However, we do not detect any point-like X-ray source running \textit{wavdetect} with a detection threshold relaxed to $10^{-5}$.

We fit the diffuse spectrum with a power-law model (including Galactic absorption) and derived a soft band flux of $f_{0.5-2 \: keV} = 1.1_{-0.3}^{+0.3} \times 10^{-15}$ erg s$^{-1}$ cm$^{-2}$. Considering the soft band flux limit for point like sources of our \textit{Chandra} observation ($f_{0.5-2 \: keV} \sim 10^{-16}$ erg s$^{-1}$ cm$^{-2}$), at least 10 unresolved X-ray sources would be required to reproduce the observed diffuse X-ray flux.

We searched in radio and optical bands for sources detected within the region of diffuse X-ray emission. In the radio observation at 1.4 GHz taken by the Very Large Array (down to a 3$\sigma$ limit of 60 $\mu$Jy/beam; \citealt{Pet03}), we found a radio lobe of a FRII galaxy (RA = 10:30:25.19, Dec = +5:24:28.50; \citealt{Pet03}) inside our region (see Figure 8, top). We also considered an archival Hubble Space Telescope WFC3 observation in the F160W filter down to mag 27 AB (bottom panel of Figure 8) and an archival 6.3 hr MUSE observation, both centered on the QSO.  The sources for which we were able to analyze MUSE spectra are marked in the image with circles of different colors (see bottom panel of Figure 8): none of them show any sign of AGN activity in their optical spectra. 
%We refer to \S 4.4 for a complete discussion on the origin of the diffuse emission and on contribution to the total X-ray emission of the sources detected in the other bands.
\begin{figure*}
\begin{center}
 \includegraphics[height=6.5cm, width=8cm, keepaspectratio]{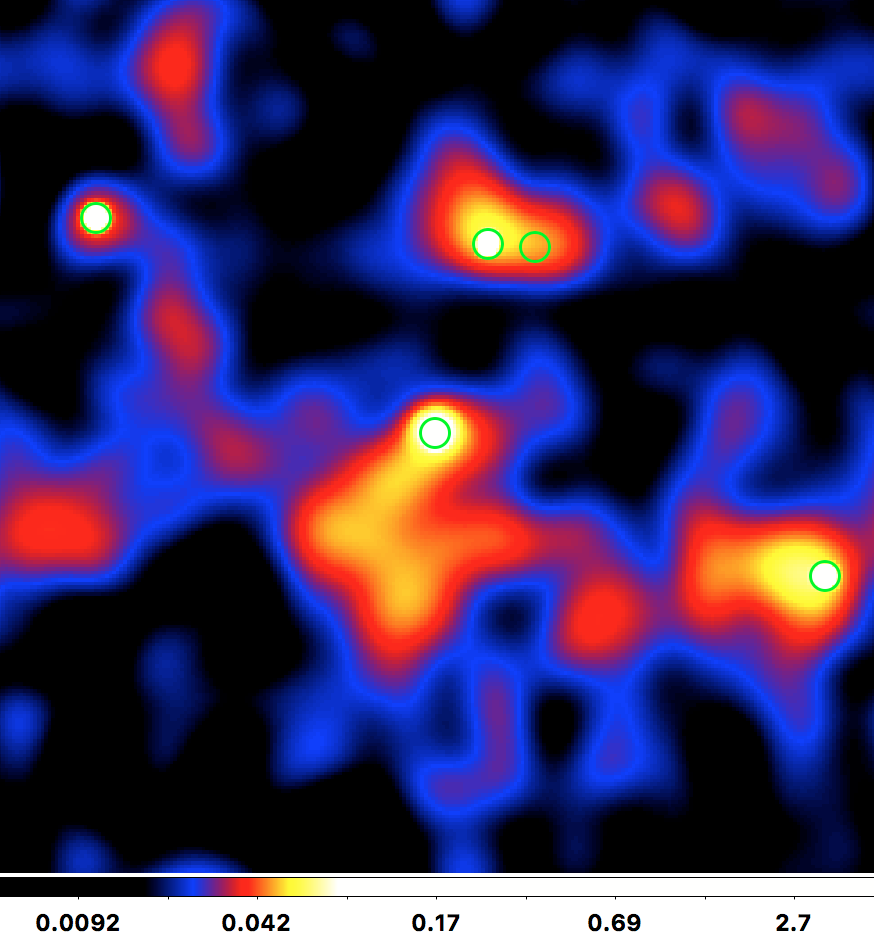}  \includegraphics[height=6.5cm, width=8cm, keepaspectratio]{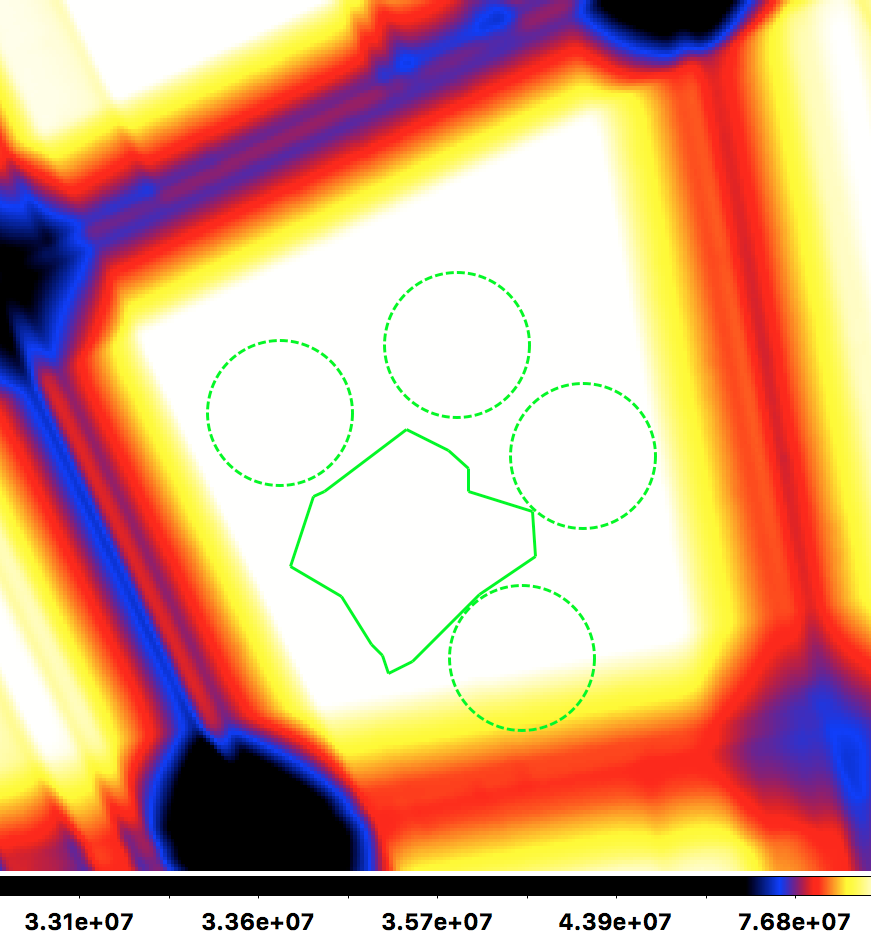}
 \includegraphics[height=6.5cm, width=8cm, keepaspectratio]{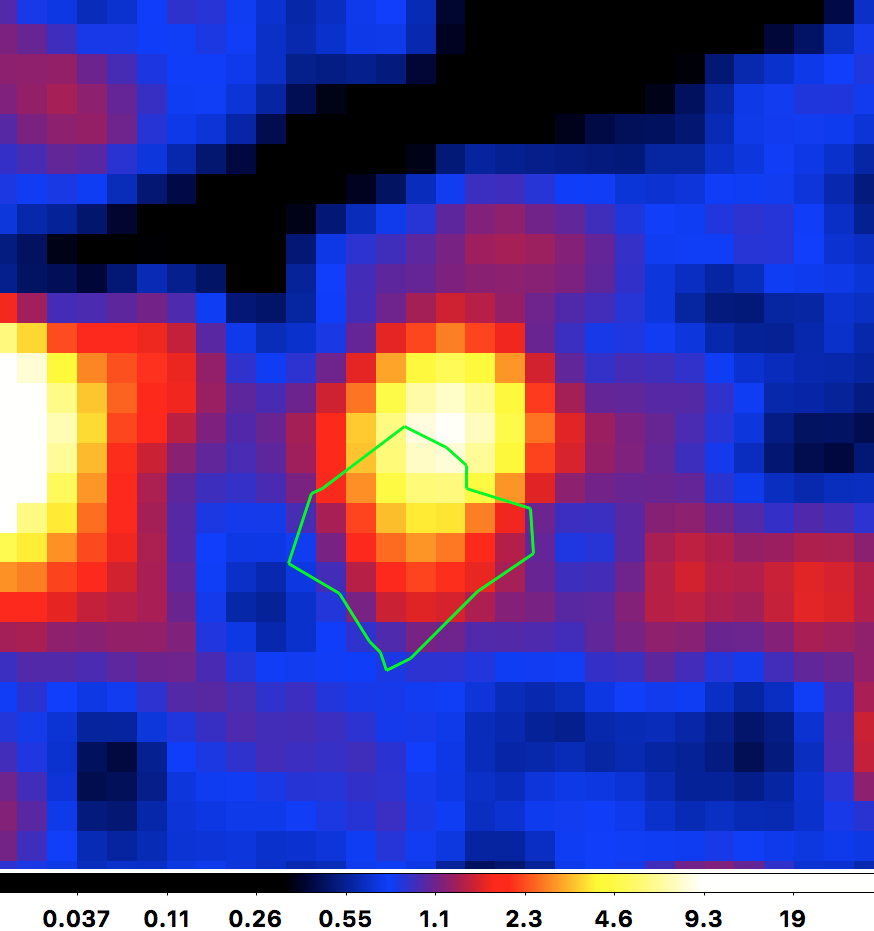}
 \caption{\textit{Left panel}: Chandra 0.5-2 keV image ($2'x2'$), smoothed with \textit{csmooth} (see text) and centered at the QSO position. North is up and East is to the left. Green circles mark point-like sources X-ray detected in the soft band. Units on the colorbar are counts per pixel. \textit{Central panel}: $2'x2'$ image of the exposure map computed at 1.4 keV and centered at the QSO position. The green 460 arcsec$^2$ region is the one used to extract the net counts of the diffuse emission southward on the QSO in the un-smoothed image, and it lays in the most sensitive peak of the exposure map. The four green dashed circles are the regions used to extract the background. Units on the colorbar are cm$^2$ s per pixel. \textit{Right panel}: XMM-\textit{Newton}-pn image in the 0.5-2 keV band ($2'x2'$), centered at the QSO position. Units on the colorbar are counts per pixel.}
 \end{center}
\end{figure*}
\begin{figure}
 \includegraphics[height=7cm, width=9cm, keepaspectratio]{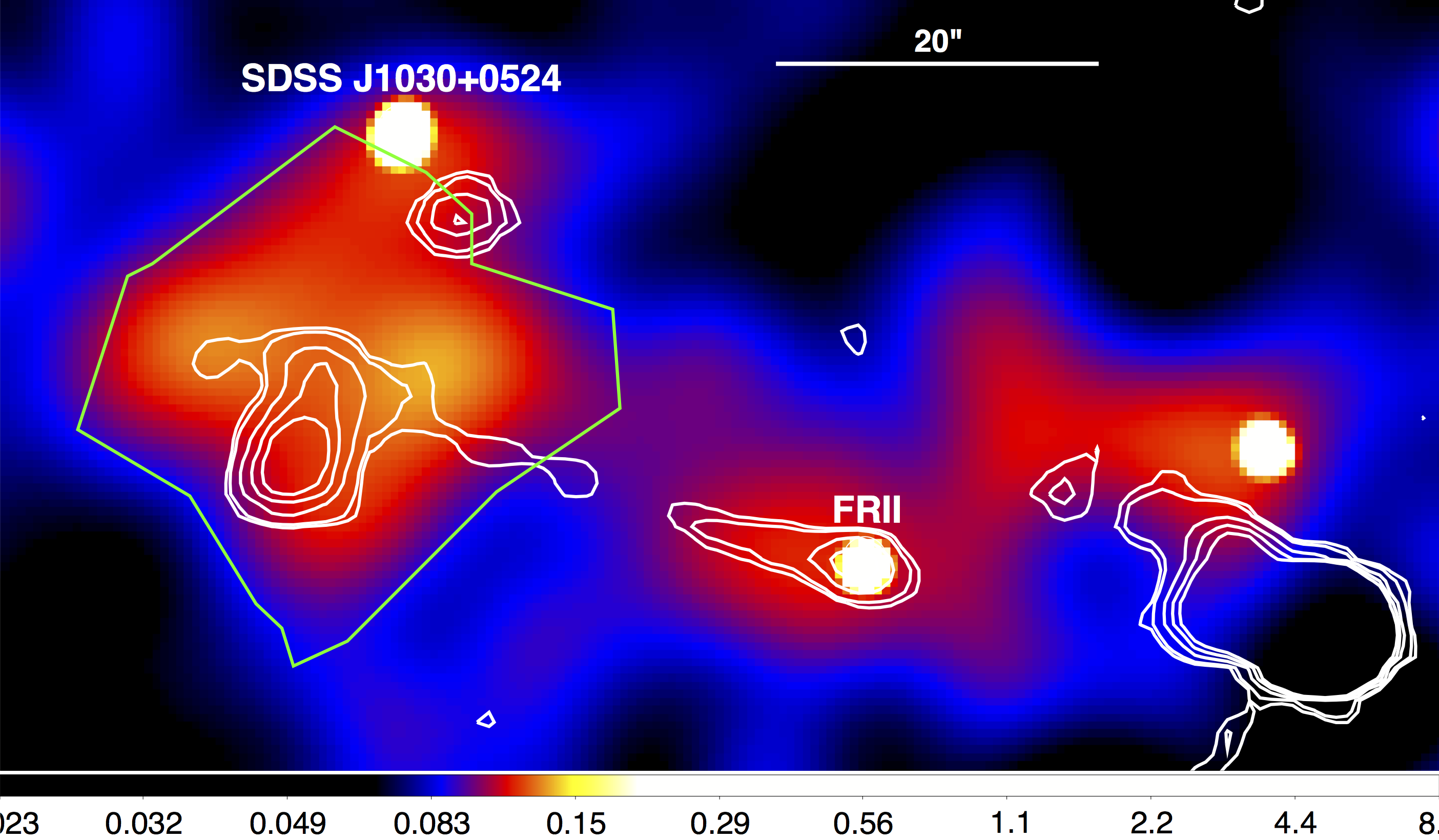}\quad \includegraphics[height=9cm, width=9cm, keepaspectratio]{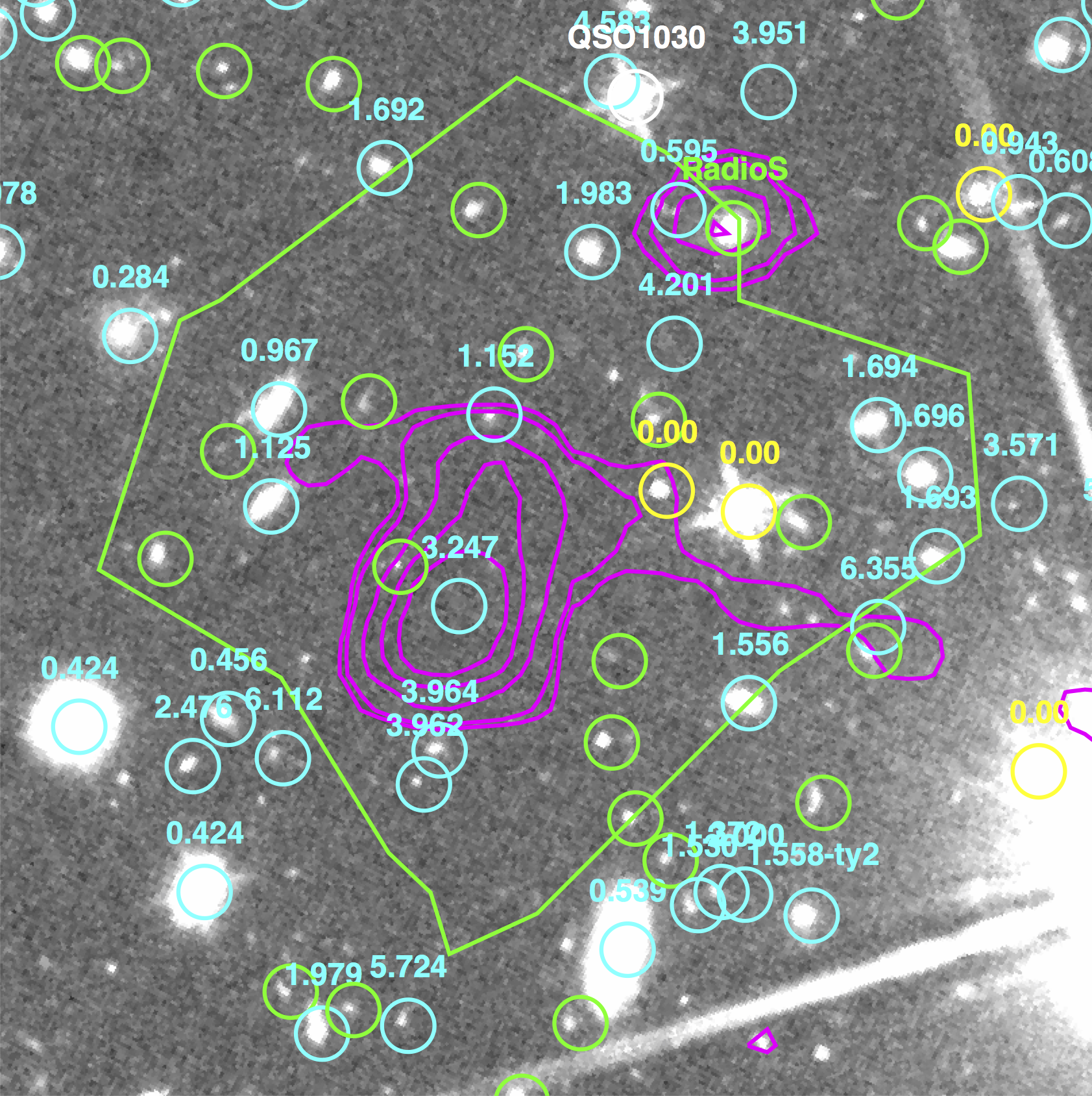}
 \caption{\textit{Top panel}:  smoothed Chandra 0.5-7 keV image of the 1.5 $\times$ 0.8 arcmin$^2$ field around SDSS~J1030+0524 and the nearby FRII radio galaxy.
Radio contours at 1.4 GHz are shown in white. Contour levels are a geometric progression in the square root of two starting at 60 $\mu$Jy. The green polygon marks the region of diffuse X-ray emission southward of the QSO.
Units on the colorbar are counts per pixel.
\textit{Bottom panel}: $0.7'\times0.7'$ HST image in the H-band. Circles are centered at the position of MUSE detected sources. The white circle marks SDSS J1030+0524. Cyan circles mark sources for which a redshift was measured, yellow circles are stars, green circles mark sources for which no redshift was measured. Magenta contours mark the emission of radio sources at 1.4 GHz. The radio lobe of a FRII radio galaxy (\citealt{Pet03}) falls within the region of X-ray diffuse emission (green polygon). In both panels, another radio source is visible at the edge of the diffuse X-ray emission (6" South-West the QSO). In all images North is up and East is to the left.}
\end{figure}

%--------------------------------------

\section{Discussion}

\subsection{Variability amplitude}

To compare the variability seen in SDSS J1030+0524 with that typically seen in AGN, we computed its normalized excess variance, as defined by \citet{Nandra97} and \citet{Tur99}, and compared it with what is measured in the samples of \citet{She17} and \citet{Pao17}.
%\begin{equation}
%	\sigma_{rms}^2 = \frac{1}{N\bar{f}^2}\sum_{i=1}^N\left[(f_i-\bar{f})^2-\sigma_i^2\right],
%\end{equation}
%where N=3, while the formal error on $\sigma^2_{rms}$ is $s_D/(\bar{f}^2\sqrt{N})$, where
%\begin{equation}
 %s_D^2 = \frac{1}{N-1}\sum_{i=1}^N\left\{\left[(f_i-\bar{f})^2-\sigma_i^2\right] -\sigma_{rms}^2\bar{f}^2\right\}^2.
%\end{equation}
%Considering the three full-band fluxes we derived in \S 3.3 for the QSO, we obtained $\sigma_{rms}^2 = 0.1 \pm 0.1$.
\citet{Pao17} measured the variability amplitude ($\sigma_{rms}^2$), in the rest-frame $2-8$ keV band, primarily for minimizing the effects of variable obscuration, of X-ray-selected AGN in the 7 Ms exposure of the \textit{Chandra} Deep Field-South (CDF-S) survey (\citealt{Luo17}). This sample includes variable and non-variable radio-quiet AGN.
\citet{She17} studied a luminous sample of four radio-quiet quasars (RQQs) at $4.10 \le z \le 4.35$, monitored by \textit{Chandra} at different epochs. 
Both \citet{She17} and \citet{Pao17} found that the X-ray variability anticorrelates with intrinsic AGN X-ray luminosity. This effect has been also observed in samples of nearby AGN and has been interpreted as the consequence of a larger BH mass in more luminous objects, which would also increase the size of the last stable orbit of the accretion disk and thus influence the overall variability produced in its innermost parts (\citealt{Pap04}).

We derived the rest-frame $2-8$ keV (observed-frame $0.3-1$ keV) band fluxes of the three observations and computed the corresponding X-ray variance and error.
In Table 3 we summarize the observed-frame full band and the rest-frame $2-8$ keV band fluxes, and luminosity obtained from our best-fit models and analysis described in \S 2.3.
\begin{table}
  \centering
  \captionsetup{justification=centering, labelsep = newline}
      \caption[]{Best-fit fluxes}
      \begin{adjustbox}{center, max width=\textwidth}
         \begin{tabular}{c c c c c}
            \hline
            \hline \rule[0.7mm]{0mm}{3.5mm}
            Observation & $\Gamma$ & f$_{(0.5-7 \: keV)}$ & f$_{(2-8 \: keV)}^{ rest}$ & L$_{(2-8 \: keV)}^{rest}$\\
           (1) & (2) & (3) & (4) & (5)\\
            \hline \rule[0.7mm]{0mm}{3.5mm}
            \textit{Chandra} 2002$^{\dagger}$ & 1.9 & 5.4$_{-2.1}^{+3.0}$ & 1.8$_{-0.7}^{+0.9}$ & 0.9$_{-0.3}^{+0.5}$\\
            \rule[0.7mm]{0mm}{3.5mm}
            XMM-\textit{Newton} 2003$^{\ddagger}$ &  2.37$_{-0.15}^{+0.16}$ & 9.78$_{-1.18}^{+0.44}$ & 5.11$_{-0.72}^{+0.33}$ & 2.80$_{-0.34}^{+0.15}$\\  
            \rule[0.7mm]{0mm}{3.5mm}
 	   \textit{Chandra} 2017 & 1.81$_{-0.18}^{+0.18}$ & 3.96$_{-0.83}^{+0.18}$ & 1.16$_{-0.37}^{+0.15}$ & 0.61$_{-0.22}^{+0.09}$\\[2pt]
            \hline 
          \end{tabular}
        \end{adjustbox}
        \begin{tablenotes}\footnotesize
		\item (1) X-ray observation of J1030+0524. (2) Photon index found or used in the fit. (3) Flux in the observed-frame $0.5-7$ keV band in units of 10$^{-15}$ erg cm$^{-2}$ s$^{-1}$. (4) Flux in the rest-frame $2-8$ keV band in units of 10$^{-15}$ erg cm$^{-2}$ s$^{-1}$. (5) Luminosity in the rest-frame $2-8$ keV band in units of $10^{45}$ erg s$^{-1}$. Errors are reported at the 1$\sigma$ level.
		\item $\dagger$ For the 2002 \textit{Chandra} observation we report the values derived from PIMMS, assuming a power-law model of $\Gamma=1.9$.
		\item $\ddagger$ For the XMM-\textit{Newton} observation we provide the photon index obtained from the joint fit and the fluxes and luminosity obtained averaging the values from the three detectors (pn, MOS1, MOS2).
	 \end{tablenotes}
\end{table}   
Considering the rest-frame $2-8$ keV band fluxes reported in Table 3, we obtained an X-ray variance $\sigma_{rms}^2 = 0.36 \pm 0.20$. The weighted mean luminosity of the three X-ray observations is $L_{2-8\: keV} = 1.23_{-0.17}^{+0.08} \times 10^{45}$ erg/s.
This value of $\sigma_{rms}^2$ is nominally 8 times higher than the average value found for QSOs of similar luminosities by \citet{Pao17} and \citet{She17}. However, because of the limited monitoring of the X-ray light curve, the formal errors on the excess variance are much smaller than the true uncertainties, 
which should be assessed with dedicated simulations (see \citealt{Pao17}). Therefore, we are not able to determine whether the observed variability is still consistent  with what is typically observed in luminous QSOs.
For instance, in 2003 XMM-\textit{Newton} may have caught the QSO in a burst period produced by an enhanced accretion episode. Further X-ray observations are needed to determine what is the typical flux state of SDSS J1030+0524, increasing the X-ray monitoring of the QSO and adding more data points to the light curve shown in Figure 6.

\subsection{Spectral variability}

The results reported in \S 3.3 highlighted a significant spectral variation between the XMM-\textit{Newton} and the \textit{Chandra} observations. 
We investigated the possibility that the high X-ray flux level and the steep QSO spectrum measured by XMM-Newton are contaminated by the diffuse X-ray emission seen southward of the QSO in the \textit{Chandra} image, which is partly included in the 15"-radius extraction region used for the analysis of the XMM-\textit{Newton} spectrum
of the QSO.
To this goal, we first considered the portion of diffuse emission falling within r < 15"  from the QSO, extracted its spectrum, and fit it with a power-law model. This spectrum contains about 1/3 of the total photons and flux of the diffuse emission reported in \S 3.4. 
Then, we fit the XMM-\textit{Newton} spectrum of the QSO with a double power-law model, where the best fit slope and normalization of one of the two power law components were
fixed to the best fit values measured for the diffuse emission within r < 15" from the QSO. As a result of this test, we found that the diffuse component contributes
less than 10\% to the QSO flux measured by XMM-\textit{Newton}, and also has negligible impact on its spectral slope. 
The contamination by the diffuse emission is therefore not able to explain the observed X-ray spectral variability.

Spectral changes are often detected in a sizable fraction of high-z AGN samples (e.g., \citealt{Pao02}), and in about 50\% of the cases such changes correlate with flux variations. In our case, the origin of the flattening of the X-ray spectral slope is unclear due to the relatively poor counting statistics that affects all the X-ray observations. This spectral variability could be related to two possible scenarios. The first one considers a change in the spectral slope related to the variation of the accretion rate with time, that makes the spectrum of the QSO steeper when the accretion rate is higher (\citealt{Sob09}). To test this scenario, we computed the X-ray Eddington ratio $\lambda_{X,E} = L_{2-10 \: keV}/L_E$, as defined in \citet{Sob09}, where $L_E = 1.3 \times 10^{38} M_{BH}/M_{\odot}$ erg s$^{-1}$ is the Eddington luminosity, and $M_{BH} = 1.4\times 10^9$ $M_{\odot}$ in SDSS J1020+0524 (\citealt{Kur07}; \citealt{DeR11}).
We computed $\lambda_{X,E}$ for the 2017 \textit{Chandra} ($\lambda_{X,E}=0.004^{+0.001}_{-0.002}$) and the 2003 XMM-\textit{Newton} ($\lambda_{X,E}=0.019^{+0.002}_{-0.001}$) observations and found that they are in general agreement with the $\Gamma - \lambda_{X,E}$ relation found by \citet{Sob09}.
The second scenario considers the "flattening" effect caused by an occultation event, for instance produced by gas clouds in the broad line region or in the clumpy torus, as sometimes observed in local AGN (\citealt{Ris07}). As shown in \S 3.2, an intervening gas cloud with N$_H \sim 5.3 \times 10^{23}$ cm$^{-2}$ is needed to reproduce the observed "flattening" of a power-law with $\Gamma \sim 2.4$, that is the photon index found in the XMM-\textit{Newton} observation. Removing the absorption term from the power-law plus absorption model (the third one reported in Table 2), we derived a full band flux of $f_{0.5-7\:keV} = 5.8^{+0.26}_{-1.22}\times10^{-15}$ erg s$^{-1}$ cm$^{-2}$, that is 1.7 times lower than the one found with XMM-\textit{Newton} (with a 3$\sigma$ significance). We conclude that the spectral and flux variabilities are not related to a simple absorption event, but that this must be coupled with an intrinsic decrease of the source power.

\subsection{The multi-wavelength SED}
In Figure 9 we provide the multi-wavelength SED of SDSS J1030+0524, which is one of the $z\sim6$ QSO best studied in different bands. 
The XMM-\textit{Newton} and \textit{Chandra} values are from this paper; LBT values from \citet{Mors14} (r, i, z bands); \textit{Spitzer} and \textit{Herschel} fluxes from \citet{Lei14} (IRAC and MIPS for \textit{Spitzer}; PACS and SPIRE for \textit{Herschel}); \textit{Scuba} values from \citet{Prid08} (at 1250, 850, 450 $\mu$m); CFHT values from \citet{Bal17} (Y, J bands); H-band and K-band values are from the MUSYC survey (\citealt{Gaw06}); VLA value from \citet{Pet03} (at 1.4 GHz); ALMA flux from \citet{Dec18} (at 252 GHz). 
The QSO SED is consistent with the combined SED of lower redshift QSOs, of \citet{Rich06} (green curve in Figure 9), showing that SDSS J1030+0524 has the typical optical properties of lower redshift luminous AGN.

\citet{Dec18} studied the FIR properties of luminous high-z AGN, based on a sample of 27 QSOs at $z \ge 5.9$ observed with ALMA. SDSS J1030+0524 is one of the few objects in the sample that is not detected in the [CII] (158 $\mu$m), and is only marginally detected in the continuum, suggesting a star formation rate (SFR) < 100 $M_{\odot}$/yr, whereas the average SFR in the sample is a few hundreds $M_{\odot}$/yr. This may suggest that SDSS J1030+0524 is in a more evolved state than the other luminous QSOs at that redshift, i.e. it may be in a stage where the star formation in its host is being quenched by its feedback (\citealt{Hop08}; \citealt{Lapi14}).

We use the full-band \textit{Chandra} flux and the 1450 \AA$\:$ magnitude of the QSO ($m_{1450 \: \AA} = 19.7$; \citealt{Ban17}) to compute the optical-X-ray power-law slope, defined as
\begin{equation}
\alpha_{ox} = \frac{log(f_{2 \: keV}/f_{2500 \: \AA})}{log(\nu_{2 \: keV}/\nu_{2500 \: \AA})},
\end{equation}
where $f_{2 \: keV}$ and $f_{2500 \: \AA}$ are the flux densities at rest-frame 2 keV and 2500 \AA, respectively. The flux density at 2500 \AA$\:$ was derived from the 1450 \AA$\:$ magnitude, assuming a UV-optical power-law slope of 0.5. We found $\alpha_{ox} = -1.76^{+0.06}_{-0.06}$, that is consistent with the mean value, $\alpha_{ox} = -1.80^{+0.02}_{-0.02}$, found for sources at the same redshift ($5.9 \le z \le6.5$: \citealt{Nan17}). The errors on $\alpha_{ox}$ were computed following the numerical method described in \S 1.7.3 of \citet{Lyo91}, taking into account the uncertainties in the X-ray counts and an uncertainty of 10\% in the 2500 \AA$\:$ flux corresponding to a mean z-magnitude error of 0.1.
Previous works have shown that there is a significant correlation between $\alpha_{ox}$ and the monochromatic $L_{2500}$ \AA$\:$ ($\alpha_{ox}$ decreases as  $L_{2500}$ \AA$\:$ increases; \citealt{Ste06}; \citealt{LR17}; \citealt{Nan17}), whereas the apparent dependence of  $\alpha_{ox}$ on redshift can be explained by a selection bias (\citealt{Zam81}; \citealt{Vig03}; \citealt{Ste06}; \citealt{She06}; \citealt{Jus07}; \citealt{Lus10}; but see also \citealt{Kel07})
The derived value is not consistent with the one found by \citet{Nan17} for XMM-\textit{Newton} data ($\alpha_{ox} = -1.60^{+0.02}_{-0.03}$), that is one of the flattest found among all $z\sim6$ QSOs. Considering also the already discussed evidence that the XMM-\textit{Newton} photon index is steeper than the mean population of QSOs at $z\sim6$, we conclude that the properties derived from XMM-\textit{Newton} data do not probably represent the typical status for SDSS J1030+0524 (that is probably more similar to that found with \textit{Chandra}), strengthening the idea that the higher flux measured by XMM-\textit{Newton} is related to an episodic burst occurred during that observation.

We also checked for the presence of long-term optical variability by comparing the J-band magnitude taken from MUSYC in 2003 (\citealt{Quad07}) with the one taken by WIRCAM in 2015 (\citealt{Bal17}). We used stars in both images to calibrate for the differences in aperture correction and in the filter response, finding a r.m.s. in the distribution of magnitude differences of $\Delta mag\sim0.04$. From 2003 to 2015 the QSO decreased its luminosity by $\Delta mag\sim0.1$. Therefore, the variation is significant only at 2$\sigma$, and if it is of the order of 10\% or less as suggested by our measurements, it would have negligible impact on the reported $\alpha_{ox}$ values.

\begin{figure*}
 \centering
 \includegraphics[height=10cm, width=15cm, keepaspectratio]{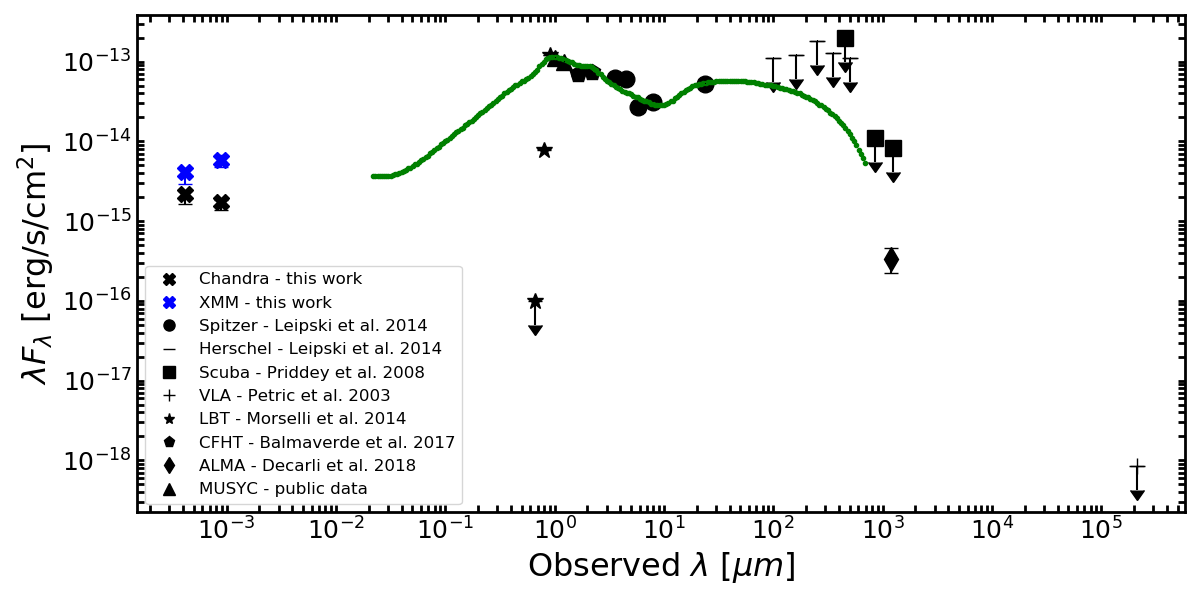}
 \caption{Multi-wavelength SED of SDSS J1030+0524. References to the points are labelled. The green curve is the combined SED for luminous lower redshift QSOs taken from \citet{Rich06}. The drop on the SED of SDSS J1030+0524 at $\lambda < 1$ $\mu m$  is produced by the Lyman alpha forest.}
\end{figure*}

\subsection{Origin of the diffuse emission}

The origin of the diffuse X-ray emission seen southward of the QSO is far from being clear. We discuss below some possible interpretations.

\subsubsection{Unresolved sources or foreground group/cluster}

In the 460 arcsec$^2$ region where we find significant excess of X-ray emission (see Figure 7, left)
we do not detect any X-ray point-like source down to $f_{0.5-2 \: keV}=10^{-16}$ erg s$^{-1}$ cm$^{-2}$ and we do not find any sign of AGN activity in any of the MUSE spectra of the optical sources 
we were able to extract in the same region. An implausibly high surface density of undetected point-like X-ray sources (100 times larger than that expected by the logN-logS relation at $f_{0.5-2 \: keV} =10^{-16}$ erg s$^{-1}$ cm$^{-2}$; \citealt{Luo17}) would be required to reproduce entirely the observed flux.

The number density of optical sources detected within the extended X-ray emission is similar to that in nearby regions, and neither the angular distribution of galaxies within the region, nor the redshifts measured with MUSE (see Figure 8, bottom) suggest the presence of a foreground group/cluster.
Therefore, we can exclude the presence of a foreground virialized group/cluster as responsible for the diffuse X-ray emission.

\subsubsection{Emission from a foreground radio galaxy}

A radio-galaxy with FRII morphology was found $\sim$40" South-West of the QSO in a relatively deep VLA observation at 1.4 GHz with $\sim$1.5" resolution (\citealt{Pet03}). 
We reanalyzed the archival VLA
data, and derived the radio contours shown in Figure 8\footnote{Our data reduction was tuned to achieve a lower resolution ($FWHM\sim3.5$") to maximize the detection efficiency of the diffuse region emission. The sensitivity limit of our image is $f_{1.4 \: GHz}\sim70$ $\mu$Jy/beam (3$\sigma$).} The nucleus of the FRII coincides with a \textit{Chandra} source detected only above 2 keV (with $\sim$30 net counts),
suggesting an extremely obscured nucleus (see Figure 8, top). The Eastern lobe of the FRII has a total radio flux of 1.7 mJy and falls within the region of diffuse emission, while the Western one is much brighter, with a total radio flux of 24 mJy. No X-ray emission is associated with the Western lobe.
A radio jet is also seen running from the radio core to the Eastern lobe. Because of the beamed nature of the jet synchrotron emission, the Eastern lobe is then
supposedly the closest to the observer. A detailed analysis of the radio source is beyond the scope of this paper. 
Here we discuss its basic properties and the likelihood that the diffuse X-ray emission seen southward
of SDSS J1030+0524 can be associated to it. 

The emission at 1.4 GHz in the Eastern lobe is not as extended as the diffuse X-ray emission. The interferometric radio
observations were, however, conducted with the "A" configuration at the VLA, which may have filtered out diffuse radio emission
on scales of tens of arcsec. As a matter of fact, low surface brightness, low significance radio emission in coincidence with the diffuse X-ray
structure and even beyond it may be present in the GMRT data of the 150 MHz TGSS survey\footnote{\url{http://tgssadr.strw.leidenuniv.nl/doku.php}}. Among the possible processes responsible for X-ray emission in radio lobes, we first investigated synchrotron models by extrapolating to the X-rays the flux densities
measured at 1.4 GHz and 150 MHz and using the spectral index $\alpha$ measured between the two radio bands ($f_\nu\propto \nu^{-\alpha}$).
Because of the widely different angular resolution between TGSS and VLA data (25" vs 3.5" $FWHM$), we also performed tests using data 
from the NVSS survey at 1.4 GHz ($\sim$45" resolution). Unfortunately, the low SNR of the TGSS data, coupled to the complex structure 
of the source (the Western lobe heavily contaminates the Eastern emission in NVSS data), prevents us from obtaing a robust estimate of the radio 
spectral index of the Eastern lobe: we measured values in the range $\alpha\sim 0.7-0.9$, depending on the adopted extraction regions.
When extrapolating the radio fluxes over more than 8 dex in frequency, this uncertainty in $\alpha$ produces a wide range of predicted X-ray emission, that can be as high as what we measured with \textit{Chandra}. Current data are therefore not sufficient to rule out this possibility, but we note, however, that it would be odd to see X-ray synchrotron emission in the Eastern but not in the Western lobe, that is 5-6 times brighter in the radio bands.

Besides synchrotron emission, there are two other possible scenarios to produce X-ray photons within a radio lobe. The first one involves Inverse Compton scattering between the relativistic electrons in the lobe and photons coming from either the cosmic microwave background (IC-CMB; e.g., \citealt{Erl06}), the synchrotron photons in the radio lobe itself (Synchro-Self Compton, SSC),
or even the photons emitted from the nucleus of the FRII (\citealt{Bru97}). The second scenario considers thermal emission produced by diffuse gas shock-heated
by the jet (\citealt{Cari02}; \citealt{Over05}). Unfortunately, we do not have enough photon statistics to distinguish between thermal and non-thermal emission based on the current low SNR X-ray data.

In the IC scenario, the observed hardness ratio of the diffuse X-ray emission would correspond to a power-law with photon index of $\Gamma=1.6\pm0.4$,
which is consistent with X-ray emission from radio lobes ascribed to IC processes (e.g., \citealt{Smail12}).
However, all the diffuse X-ray emission is associated to the fainter,
Eastern radio lobe, whereas, if it was produced by SSC or IC, X-ray emission in the Western lobe, which is > 6 times more powerful in the radio band, would be expected as well. In the case of anisotropic scattering of photons from the FRII nucleus, backward scattering would be favored (\citealt{Bru97})
and one would then expect the farthest (Western) lobe to be the brightest, which is not the case. Ascribing the observed diffuse X-ray emission
to processes associated with the FRII radio lobe is not therefore entirely convincing.

As an alternative, the diffuse X-ray emission may be associated with shocked gas, as is sometimes seen in distant ($z\sim 2$)
radio galaxies that are embedded in gas-rich large scale structures (\citealt{Over05}). As there is no spectroscopic redshift for the host of the FRII,
we first derived a K-band magnitude from the public data of the MUSYC survey (\citealt{Gaw06}), and then estimated a redshift of $z\approx 1-2$,
based on the K-band magnitude vs redshift relation observed for radio-galaxies (e.g., \citealt{Wil03}).
Assuming that the X-ray emission is thermal, and that both the FRII radio galaxy and the diffuse gas are embedded in a large scale structure at $z = 1.7$ 
(there is a indeed weak evidence for a spike at $z = 1.69$ in the redshift distribution of MUSE sources, see the also bottom panel of Figure 8), we derived a temperature of
$T\gtrsim 4$ keV for the X-ray emitting gas (using the {\tt apec} model within XSPEC with $0.3 \times$ solar metal abundances).
Some further diffuse X-ray emission can be recognized in Figure 8 (top) in correspondence of the FRII radio jet and North-West of the FRII core. This emission is
at very low SNR, but, if real, it may suggest the presence of other diffuse hot gas in a putative large scale structure.
At this redshift, the extension of the diffuse X-ray emission southward of SDSSJ1030+0524 would correspond to 240 physical kpc,
and its luminosity to $L_{2-10 \: keV} = 3\times 10^{43}$ erg s$^{-1}$.
Assuming that the point-like X-ray emission observed in the core of the FRII traces the accretion luminosity, we can estimate the power carried out by the jet
towards the lobe by considering that this is generally equal or larger than the accretion luminosity (\citealt{Ghis14}).
At $z=1.7$, using a photon index of $\Gamma=1.8$, the hard X-ray source at the FRII nucleus would be a Compton-thick AGN ($N_H\approx1.5\times 10^{24}\;cm^{-2}$)
with rest-frame deabsorbed 2-10 keV luminosity of $L^{rest}_{2-10}\sim 10^{44}$ erg/s. Adopting a bolometric correction of 30, as appropriate for these
X-ray luminosities (e.g., \citealt{Lus11}), we estimate a total accretion luminosity, and hence a total jet power of $P_{jet} \gtrsim 3\times 10^{45}$ erg s$^{-1}$.
From the fit to the diffuse X-ray emission we derived a gas density of $n\sim 4\times10^{-3}$ cm$^{-3}$, and hence a total thermal energy of
$E_{th}\sim nVkT\sim 5\times 10^{60}$ erg, assuming a spherical volume of radius 120 kpc. To deposit such amount of energy in the gas, the jet would
have had to be active at that power for at least 100 Myr (even assuming that the 100\% of the jet power is transferred to the gas),
which is larger than the typical lifetime of FRII jets ($\sim1.5\times 10^7$ yr; \citealt{Bird08}).
Because of the many uncertainties and assumptions, the above computation must be taken with caution.
However, it shows that even thermal emission from gas shock-heated by the FRII jet is not a secure interpretation.
 
\subsubsection{QSO feedback and X-ray jets at $z = 6.31$}

Both analytical/numerical models and simulations of early BH formation and growth postulate that a non-negligible fraction of the energy released
by early QSOs can couple with the surrounding medium producing significant feedback effects on it (\citealt{Dub13}; \citealt{Cos14}; \citealt{Bar17}; \citealt{Gill17}). In this scenario, the diffuse X-ray emission may be related to the thermal cooling of environmental gas shock-heated by QSO outflows.
This gas can be heated to temperatures higher than $10^8$ K on scales that may extend well beyond the virial radius of the dark matter halo hosting the QSO,
and reach hundreds of kpc from the QSO depending on the gas density and host halo mass (e.g., \citealt{Gill17}). Significant X-ray emission in the X-ray band
is then expected (\citealt{Cos14}). Also, the morphology of the hot gas may be highly asymmetric, depending on the outflow opening angle (\citealt{Bar17})
and even be unipolar, depending on the gas distribution in the BH vicinity (\citealt{Gab13}). The morphology of the diffuse X-ray emission suggests
that at least part of it may be indeed associated with the QSO. In fact, a "bridge" of soft X-ray emission appears to originate from the QSO and extend
into the South-Eastern part of the diffuse X-ray structure (see Figure 7, left-panel).

If the observed diffuse X-ray emission is interpreted as thermally emitting gas at $z=6.3$ (we used again the {\tt apec} model within XSPEC with 0.3$\times$ solar
metal abundances), then this should have a temperature of $T\gtrsim 10$ keV, and extend asymmetrically for about 150 physical kpc from the QSO.
This is consistent with the simulations above. The observed X-ray emission would then correspond to a luminosity of $L^{rest}_{2-10 \: keV} = 5\times 10^{44}$ erg s$^{-1}$.
As above, we computed the thermal energy of the X-ray emitting gas by assuming that it is distributed in a sphere of 75 kpc radius. We derived
a total thermal energy of $\approx 10^{61}$ erg. This is within a factor of two consistent with the predictions of \citet{Gill17}. In that paper it was
calculated that an accreting BH growing to $10^9$ $M_{\odot}$ by $z=6$, such as that observed in SDSS J1030+0524, may deposit $\sim5 \times 10^{60}$ erg of
energy in the surrounding medium through continuous, gas outflows. Furthermore, based on the thermal model fit we obtain a total gas mass
of $M_{gas}\sim 1.2\times 10^{12}\; M_{\odot}$ and hence a total dark matter halo mass of $\ge8\times 10^{12}\; M_{\odot}$,
which would be consistent with the idea that early luminous QSOs form in highest peaks of the density field in the Universe,
as further supported by the candidate galaxy overdensity measured around SDSS J1030+0524 (\citealt{Mors14}; \citealt{Bal17}).
Again, we note that many caveats apply that are related to our assumptions
and uncertainties in the physical parameters derived from low SNR X-ray data, so that the above conclusions are still speculative.

%By considering that the bolometric luminosity of SDSS J1030+0524 is $L_{bol} \sim 10^{47}$
%erg s$^{-1}$ [check precise number], less than 1\% of the AGN power would be needed to heat the surrounding gas and produce the observed  X-ray luminosity.

The QSO may also be responsible for the extended X-ray emission through non-thermal radiation mechanisms. In particular, it has been proposed that the emission
of jets and lobes in high redshift QSOs may be best probed in the X-rays rather than in the radio band. This is because the energy density of the CMB increases
as $(1+z)^4$, causing inverse Compton scattering to dominate over synchrotron emission the energy losses of relativistic electrons (\citealt{Ghis14}; \citealt{Fab14}). Despite the large uncertainties arising from the low photon statistics
and from the image smoothing process, the bridge of soft X-ray emission originating from the QSO in the Eastern part of the diffuse structure
reminds of an X-ray jet that is possibly powering a diffuse structure (X-ray lobe). Following \citet{Fab14} and \citet{Ghis14},
we considered a simple jet IC-CMB model in which the QSO produces a relativistic jet with power equal to the accretion power ($10^{47}$ erg s$^{-1}$).
We assumed that: i) the total power of the relativistic electrons injected in the lobe 
is 10\% of the jet power; ii) the magnetic field strength in the lobe is $15\; \mu G$ (near equipartition is assumed between magnetic field and relativistic particles);
iii) the lobe is a sphere of $\approx$50 kpc radius. A fiducial power-law spectrum is also assumed for the injected electrons (see \citealt{Ghis14}).
With the above assumptions, both the X-ray luminosity and hardness ratio of the diffuse emission can be reproduced by the IC scattering of CMB photons
by the electrons in the lobe, which in turn may also contribute to some of the observed radio emission through synchrotron radiation. The above computation, despite being admittedly uncertain and relying on strong assumptions, provides another plausible emission
mechanism for the observed diffuse X-ray emission.

%--------------------------------------

\section{Summary and conclusions}

In this paper we have reported on the $\sim$500 ks \textit{Chandra} observation of the QSO SDSS J1030+0524. This is the deepest X-ray observation ever achieved for a $z\sim6$ QSO.
Our main results are the following:

\begin{itemize}
\item The QSO has been detected with $\sim$125 net counts in the full band with no evidence of either significant spectral or flux variability during the \textit{Chandra} observations. The spectrum is well fit by a single power-law with $\Gamma=1.81_{-0.18}^{+0.18}$, that is consistent with the mean value found for luminous AGN at any redshift. No evidence is found for significant absorption (N$_H = 4.6^{+2.7}_{-4.6} \times 10^{23}$ cm$^{-2}$), nor for other additional spectral features.\\

\item A comparison between the QSO X-ray spectral properties in our \textit{Chandra} data with those obtained from a past XMM-\textit{Newton} observation (\citealt{Far04}; \citealt{Nan17}) revealed that the QSO significantly varied. The full band flux decreased by a factor of 2.5 from the XMM-\textit{Newton} to the \textit{Chandra} observations while the spectrum became flatter ($\Delta \Gamma \approx -0.6 $). We verified that these variations are not related to calibration issues. We discussed the possibility that the hardening of the spectral slope is intrinsic and related to variations of the accretion rate. As an alternative, a variation of the obscuration level along the line of site (with $\Gamma$ fixed at the XMM-\textit{Newton} value) is not sufficient to explain alone the observed variations. However, because of the limited monitoring of the X-ray light curve and the poor counting statistics, we were not able to disentangle between the different scenarios.\\

\item We provided the SED of SDSS J1030+0524, that is one of the best sampled for a $z>6$ QSO. The SED is consistent with the mean SED of luminous AGN at lower redshift, but it differs in the FIR and sub-millimeter bands with that found for other QSOs at $z\sim6$. This difference may suggest that SDSS J1030+0524 is in a more evolved state (i.e., with quenched SFR) than the other luminous QSOs at that redshift. We also computed the optical-X-ray power-law slope for the \textit{Chandra} observation, finding $\alpha_{ox} = -1.76^{+0.06}_{-0.06}$.
Comparisons between the $\alpha_{ox}$ and the photon index found by XMM-\textit{Newton} with those found by \textit{Chandra}, suggest that the properties derived from XMM-\textit{Newton} data do not probably represent the typical status for SDSS J1030+0524, strengthening the idea that the higher flux measured by XMM-\textit{Newton} is related to an episodic burst occurred during the XMM-\textit{Newton} observation.\\

\item We detected significant diffuse X-ray emission that extends for 30"x20" southward the QSO, with a SNR = 5.9, hardness ratio of $HR=0.03_{-0.25}^{+0.20}$, and soft band flux of $f_{0.5-2 \: keV}= 1.1_{-0.3}^{+0.3} \times 10^{-15}$ erg s$^{-1}$ cm$^{-2}$. We verified that an implausibly high surface density of undetected point-like X-ray sources would be required to reproduce entirely the observed flux. Based on HST and MUSE date, we also excluded the presence of foreground a galaxy group/cluster.  We discussed different scenarios for the origin of this diffuse emission. The first scenario considers the contribution of a radio lobe from a foreground FRII source that could emit X-ray photons via synchrotron or inverse Compton processes.
The large uncertainties on the radio measurements and the low X-ray statistics prevented us from excluding or confirming either possibility. However, the absence of X-ray emission from the brightest radio lobe poses challenges to any scenario involving non-thermal processes, as we would expect stronger X-ray emission in the other lobe, which is > 6 times more powerful in the radio band but has no X-ray diffuse emission. Alternatively, as high-z radio-galaxies are often found in non-virialized large scale structures, the diffuse X-ray may arise from a reservoir of gas in this structure shock-heated by the FRII jet.

In the second scenario, the diffuse X-ray emission may probe the feedback produced by SDSS J1030+0524 on its close environment. In that case, if the diffuse emission is thermal, the gas should have a temperature of $T\ge10$ keV, and extend asymmetrically for about 150 physical kpc from the QSO, in agreement with simulations of early BH formation. In addition, supposing that SDSS J1030+0524 is producing a relativistic jet, this would be best probed in the X-rays rather than in the radio band, as the electron energy losses would be dominated by IC scattering of the strong CMB photon field, rather than by synchrotron emission. The energetics, scales and spectral hardness of the observed X-ray emission would also be consistent with this interpretation.\\

\end{itemize}
We conclude that SDSS J1030+0524 is one of the best objects to study the spectral properties and the environment of high redshift AGN. New X-ray observations are needed to check the QSO light-curve and to constrain the origin of the extended emission via spectral analysis. In particular, for the next future \textit{Chandra} and XMM-\textit{Newton} monitoring of SDSS J1030+0524 would provide additional information on the QSO variability and the origin of the diffuse emission seen southward the QSO. This will greatly help developing the science cases for future X-ray missions, such as \textit{Athena} and \textit{Lynx}, that will shed new light on the high redshift frontier.

\begin{acknowledgements}

We acknowledge useful discussion with S. Ettori, G. Brunetti, G. Ghisellini, R. Decarli, E. Torresi, E. Lusso, S. Gallerani.
We thank the anonymous referee for his/her useful comments and suggestions that improved the quality of this work.
We acknowledge financial contribution from the agreement ASI-INAF n.2017-14-H.O and ASI-INAF I/037/12/0.
The scientific results reported in this article are based on observations made by the \textit{Chandra} X-ray Observatory.
This work made use of data taken under the ESO program ID 095.A-0714(A).
FC acknowledges funding from the INAF PRIN-SKA 2017 program 1.05.01.88.04.

 \end{acknowledgements}

\bibliographystyle{aa}
 \bibliography{riccardo}

\end{document}